\documentclass[conference]{IEEEtran}

\IEEEoverridecommandlockouts
\usepackage{amsmath,amssymb,amsfonts}
\usepackage{algorithmic}
\usepackage{graphicx}
\usepackage{subcaption}
\usepackage[margin=1in]{geometry}
\usepackage{textcomp}
\usepackage{multirow}
\usepackage[table,xcdraw]{xcolor}
\usepackage{tikz}
\usetikzlibrary{shapes,arrows.meta,positioning}
\usepackage{circuitikz}
\usepackage{csquotes}
\usepackage{pgfplots}
\usepackage{float}
\usepackage{subcaption}
\usepackage{nicefrac}
\usepackage{adjustbox}
\usepackage{siunitx}
\usepackage{caption}
\usepackage{algorithm}
\usepackage{multirow}
\usepackage{hyperref}

\usepackage{scalerel}
\usepackage{xspace}

\usetikzlibrary{svg.path}
\definecolor{orcidlogocol}{HTML}{A6CE39}
\tikzset{
  orcidlogo/.pic={
    \fill[orcidlogocol] svg{M256,128c0,70.7-57.3,128-128,128C57.3,256,0,198.7,0,128C0,57.3,57.3,0,128,0C198.7,0,256,57.3,256,128z};
    \fill[white] svg{M86.3,186.2H70.9V79.1h15.4v48.4V186.2z}
                 svg{M108.9,79.1h41.6c39.6,0,57,28.3,57,53.6c0,27.5-21.5,53.6-56.8,53.6h-41.8V79.1z M124.3,172.4h24.5c34.9,0,42.9-26.5,42.9-39.7c0-21.5-13.7-39.7-43.7-39.7h-23.7V172.4z}
                 svg{M88.7,56.8c0,5.5-4.5,10.1-10.1,10.1c-5.6,0-10.1-4.6-10.1-10.1c0-5.6,4.5-10.1,10.1-10.1C84.2,46.7,88.7,51.3,88.7,56.8z};
  }
}

\newcommand\orcidicon[1]{\href{https://orcid.org/#1}{\mbox{\scalerel*{
\begin{tikzpicture}[yscale=-1,transform shape]
\pic{orcidlogo};
\end{tikzpicture}
}{|}}}}

\usepackage{soul}
\def\BibTeX{{\rm B\kern-.05em{\sc i\kern-.025em b}\kern-.08em
    T\kern-.1667em\lower.7ex\hbox{E}\kern-.125emX}}
\begin{document}

\makeatletter
\newcommand{\linebreakand}{
  \end{@IEEEauthorhalign}
  \hfill\mbox{}\par
  \mbox{}\hfill
  \begin{@IEEEauthorhalign}
}
\makeatother

\title{Stealth by Conformity: Evading Robust Aggregation through Adaptive Poisoning \thanks{This PhD project receives funding from the Cyber AI Hub Doctoral Training Program at CSIT, Queen’s University Belfast, supported by the UK Government as part of the New Deal for Northern Ireland, administered through Innovate UK/UKRI.}}

\author{\IEEEauthorblockN{Ryan McGaughey\IEEEauthorrefmark{1},
Jes\'us Mart\'inez del Rinc\'on\IEEEauthorrefmark{2} and
Ihsen Alouani\IEEEauthorrefmark{3}}
\IEEEauthorblockA{Centre for Secure Information Technologies (CSIT),
Queen's University Belfast, UK\\
Email: \IEEEauthorrefmark{1}rmcgaughey03@qub.ac.uk \orcidicon{0009-0006-2823-8522},
\IEEEauthorrefmark{2}j.martinez-del-rincon@qub.ac.uk \orcidicon{0000-0002-9574-4138},
\IEEEauthorrefmark{3}i.alouani@qub.ac.uk \orcidicon{0000-0001-5102-8087}}}

\maketitle
\newcommand{\ihsen}[1]{\textcolor{blue}{ \@ \textit{Ihsen}: #1}}
\newcommand{\ourapproach}{\textsc{ChamP}\xspace}
\newcommand{\ourapproachfull}{Chameleon Poisoning (\textsc{ChamP})\xspace}
\newcommand{\bsc}{\textsc{}\xspace}

\begin{abstract}
Federated Learning (FL) is a distributed learning paradigm designed to address privacy concerns. However, FL is vulnerable to poisoning attacks, where Byzantine clients compromise the integrity of the global model by submitting malicious updates. Robust aggregation methods have been widely adopted to mitigate such threats, relying on the core assumption that malicious updates are inherently out-of-distribution and can therefore be identified and excluded before aggregating client updates.

In this paper, we challenge this underlying assumption by showing that a model can be poisoned while keeping malicious updates within the main distribution. We propose \ourapproachfull, an adaptive and evasive poisoning strategy that exploits side-channel feedback from the aggregation process to guide the attack.
Specifically, the adversary continuously infers whether its malicious contribution has been incorporated into the global model and adapts accordingly. This enables a dynamic adjustment of the local loss function, balancing a malicious component with a camouflaging component, thereby increasing the effectiveness of the poisoning while evading robust aggregation defenses.

\ourapproach enables more effective and evasive poisoning, highlighting a fundamental limitation of existing robust aggregation defenses and underscoring the need for new strategies to secure federated learning against sophisticated adversaries.
Our approach is evaluated in two dataset reaching an average increase of 47.07\% in attack success rate against nine robust aggregation defenses.
\end{abstract}


\section{Introduction}\label{Introduction}
Federated Learning (FL) was introduced in \cite{mcmahan_communication-efficient_2017} as a distributed training paradigm designed to enhance privacy by keeping data on client devices. Instead of sharing raw data, only model parameters are exchanged, thereby mitigating many of the privacy risks inherent to centralized learning. This makes FL especially valuable in data-sensitive domains, where client data must remain local and protected against potential leakage.

Under this framework, in each round of FL training, each client trains a local model on its private dataset, then transmits the model updates (e.g., gradients or weights) to a central server, which aggregates these updates—typically via weighted averaging based on dataset size— to produce a refined global model, which is then redistributed to clients.

In the original FL setup proposed in \cite{mcmahan_communication-efficient_2017}, FedAvg was proposed as the aggregation scheme. This aggregation was done by taking the parameter by parameter average update to the server by each client.

 \begin{figure}[tp]
    \centering
    \includegraphics[width=\linewidth]{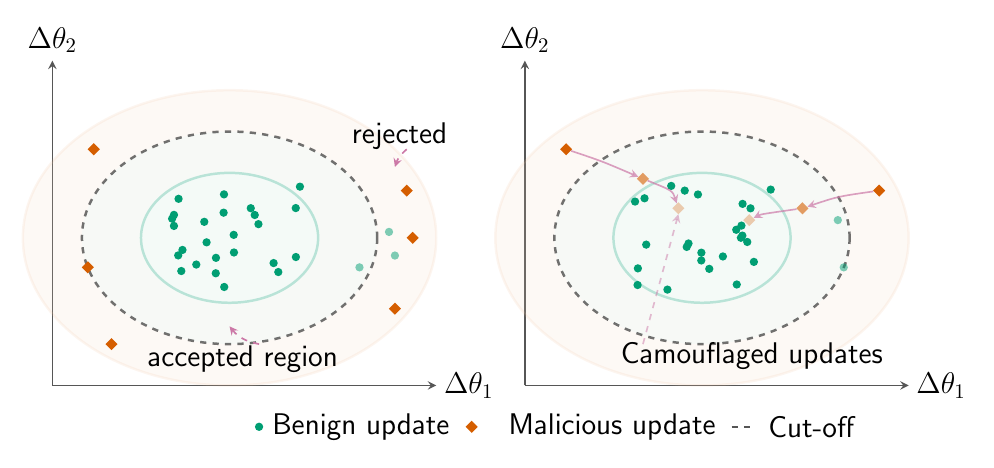} 
    \caption{Conceptual description of our adaptive poisoning attack against a RA scheme. Data points represent instances of model updates $\Delta\theta$. }
    \label{fig:overview}
\end{figure}

While FL mitigates many traditional privacy risks and provides a scalable alternative to centralized learning, its decentralized nature also introduces new vulnerabilities. Since the central server has no direct access to client data, it must trust the updates provided by participating clients. This exposes the system to poisoning attacks, where malicious or Byzantine clients deliberately craft updates to manipulate the global model \cite{bagdasaryan_how_2020}. Unlike random noise or benign irregularities, these adversarial updates are designed to degrade model performance, introduce targeted misclassifications, or embed hidden backdoors; all while evading detection \cite{fang_local_2020,shejwalkar_manipulating_2021,xie_dba_2019}.

To mitigate the risk of poisoning, a prominent line of defense in FL has been the development of Robust Aggregation (RA) methods. The core rationale of these methods is that malicious updates can be distinguished from benign ones because they are expected to deviate significantly from the majority. By assuming that most clients behave honestly and that adversarial contributions are statistical outliers, RA seeks to preserve the integrity of the global model through anomaly-resistant techniques such as trimmed mean, coordinate-wise median, or geometric approaches \cite{blanchard_machine_2017,mhamdi_hidden_2018,yin_byzantine-robust_2018}. In practice, these methods attempt to filter or exclude suspicious updates before aggregation, thereby preventing Byzantine clients from disproportionately influencing the model.

This approach, however, relies on a critical assumption: that malicious updates are inherently out-of-distribution relative to benign ones. In this paper, we challenge that assumption, showing that poisoning can remain both effective and statistically consistent with benign updates. Our rationale is twofold. First, modern FL operates in very high-dimensional parameter spaces with heterogeneous clients, so benign updates naturally exhibit broad, anisotropic variability. Common robust aggregators, such as coordinate-wise trimming/median, geometric median, Krum variants, instantiate “out-of-distribution” via deviation rules whose acceptance regions widen along directions of natural variance. We exploit this by designing an adaptive, camouflaged poisoning method that shapes its updates to remain within typical ranges, norms, and angular similarities, while subtly steering the global model in task-critical directions. Second, deep networks have steep, non-convex loss landscapes: small, well-placed parameter shifts can induce disproportionately large behavioral changes. From this perspective, our problem formulation is the dual of adversarial examples problem. Whereas adversarial attacks constrain perturbations (e.g., by projecting into an $\ell_p$-norm ball in the input space) our method operates in the parameter-update space, keeping updates within the robust aggregator’s “acceptable” range as implied by honest variability and deviation thresholds. The adversarial updates dual problem is outlined conceptually in Fig. \ref{fig:overview}. This, however, raises a key challenge: \textbf{how can we reliably steer updates into the acceptable region without white-box access to the server or its aggregation rule?}

To address this challenge, we extract side-channel feedback from the aggregated model. After each round, the updated global model is probed and a targeted membership-inference test is run by the malicious client on the poisoned/backdoor set to estimate whether—and to what extent—the malicious contribution has been incorporated by the aggregator. The resulting signal —binary or graded— drives an adaptive regularization term that updates the camouflage component of the local objective, dynamically rebalancing attack strength against distributional conformity. This feedback mechanism enables black-box steering of updates into the aggregator’s acceptance region while sustaining poisoning efficacy across rounds.

We evaluate our attack on nine state-of-the-art RA methods from different categories, including coordinate-wise, distance-based, and clustering-based approaches, and show that all of them fail to provide robustness against our approach. Our experiments demonstrate that robust aggregation schemes remain vulnerable to adaptive adversaries, challenging the prevailing assumption that robust aggregation ensures resilience, and redefining the threat model for secure FL.

The main contributions of this paper are as follows:
\begin{itemize}
    \item We propose a new attack strategy that adaptively camouflages malicious updates to evade detection. The attack dynamically balances a malicious objective with a camouflage loss to keep updates within the acceptance region of robust aggregators. 
    \item The camouflage loss regularization term is updated continuously based on a side-channel information that we collect from the aggregated model. We introduce a black-box feedback mechanism that estimates whether the adversary's contribution has been incorporated into the aggregated model. This signal is extracted through membership inference on the poison/backdoor data and is used to adaptively regulate the camouflage loss. 
    \item We evaluate the attack across multiple datasets, model architectures, and robust aggregation techniques. Results show that our method consistently evades robust aggregation, achieving high poisoning success rates while maintaining statistical conformity with benign updates.
\end{itemize}

\section{Background}\label{Background}
\subsection{Federated Learning}
In the classic FL setup, we have a central server coordinating with $N$ clients, denoted as $C_n$, where $n \in {1, 2, \dots, N}$. The goal of FL, assuming all benign clients, is to minimize the global objective loss function $\mathcal{L}$ at a given training round $t \in {1, 2, \dots, T}$, where $T$ is the total number of training rounds, to create a final global model $G^T$. For a global training round $t$, the benign clients aim to minimize:
\begin{equation*}
    \mathcal{L}(G^t) = \frac{1}{N}\sum^N_{n=1}\mathcal{L}_n(g_n^t)
\end{equation*}
where $\mathcal{L}_n$ and $g_n$ are the local objective loss function and the local model for client $C_n$, defined over its local dataset $D_n = \{(x, y) \mid x \in \mathcal{X},\ y \in \mathcal{Y} \}$. To optimize this objective, the following set of steps is performed at every training round $t$:
\begin{enumerate}
    \item \textbf{Broadcasting of the Global Model}: The central server broadcasts the global model ($G^{t-1}$) to a subset of local clients.
    \item \textbf{Local Training}: Each local client $C_n$ selected by the central server in the aggregation round performs local training on it's private data-set $D_n$ to produce a new local model $g_n^{t}$.
    \item \textbf{Aggregation}: The central server collects the resultant local models and aggregated them in accordance to an aggregation scheme $G^{t} = Agg\{g_n^{t} :n \in {1, 2, \dots, N}\}$.
\end{enumerate}
These steps are repeated until a given convergence condition is met. In the classic FL setup, the aggregation scheme used is FedAvg, the aggregation is outlined for a given training round $t$ below for comparison in our discussion on robust aggregation schemes. We assume that every client is training using the same model architecture, and say for a model with $K$ co-ordinates, that for a co-ordinate $k$ we have:
\begin{align}
    \text{FedAvg}\{g_{nk}^t : n \in N, k \in K\} 
    &= \frac{1}{N}\sum_{n=1}^N g^t_{n,k}
\end{align}
which is the average of every parameter in each local client update to the central server.
\subsection{Backdoor Attacks in FL}
In our setting, we assume that we have $M \subseteq N$ malicious client(s) in the FL system, that aim to poison the global model $G$, to learn a malicious secondary objective as well as the the primary objective.. Therefore, our system consists of $N$ clients where we say, for simplicity, that we have up to $C_M$ malicious clients present. The malicious clients attempt to train their local models to be poisonous in order to affect the global model enough for their desired objective to be realised. The aim of the attacker(s) could be to completely destroying the behaviour of $G$ across specific samples, or tricking the global model to behave a specific way when ran across a specific subset of data points.

In this paper we implement backdoor attacks \cite{gu_badnets_2019}, where a malicious actor attempts to impact the global model by training it's local model to misclassify specific samples when a backdoor trigger is seen. That includes altering not only the labels of the samples, but also the data itself to include a specific trigger implemented by the malicious client\footnote{Traditionally, semantic samples are discussed in \cite{bagdasaryan_how_2020} however they will not be discussed here.}. We define any clients local dataset $D_n$ as follows:
\begin{equation}
D_n = \{x_i,y_i\}, \text{ where } i \in \{1,...,|D_n|\}
\end{equation}
where $y_i$ represents the class of data sample $x_i$, which has a value in the set $\mathcal{Y}$ and $x_i$ is a data point in the set $\mathcal{X}$. There are two main classificatons of backdoor attacks, targeted attacks, where samples of a source class $y^s$ are altered to be of source class $y^\tau$, and untargeted where source class samples are altered to be of any source class that is not $y^s$. So, for a malicious client $C_m$, their local dataset is altered during a targeted backdoor attack such that:
\begin{equation}
\hat{D}_m = \{(\hat{x}_i,\hat{y}_i) | (x_i,y_i)\} \in D_m, \hat{y}_i = 
\begin{cases}
  y^\tau_i & \text{if } y_i=y^s \\
  y_i & \text{otherwise}
\end{cases}
\end{equation}
For untargeted attacks, we have a very similar formula, except the class of the samples is denoted as any class which is not $y^s$:
\begin{equation}
\hat{D}_m = \{(\hat{x}_i,\hat{y}_i) | (x_i,y_i)\} \in D_m, \hat{y}_i = 
\begin{cases}
  \text{any } y \in \mathcal{Y} \setminus\{y^s\}\\
   \text{if } y_i=y^s \\
 \\
  y_i  \text{ otherwise}
\end{cases}
\end{equation}
In this paper, we focus on targeted attacks, which we outline visually in Fig.\ref{fig:backdoor}.

\begin{figure}
    \centering
    \includegraphics[width=\linewidth,trim=72 80 30 15,clip]{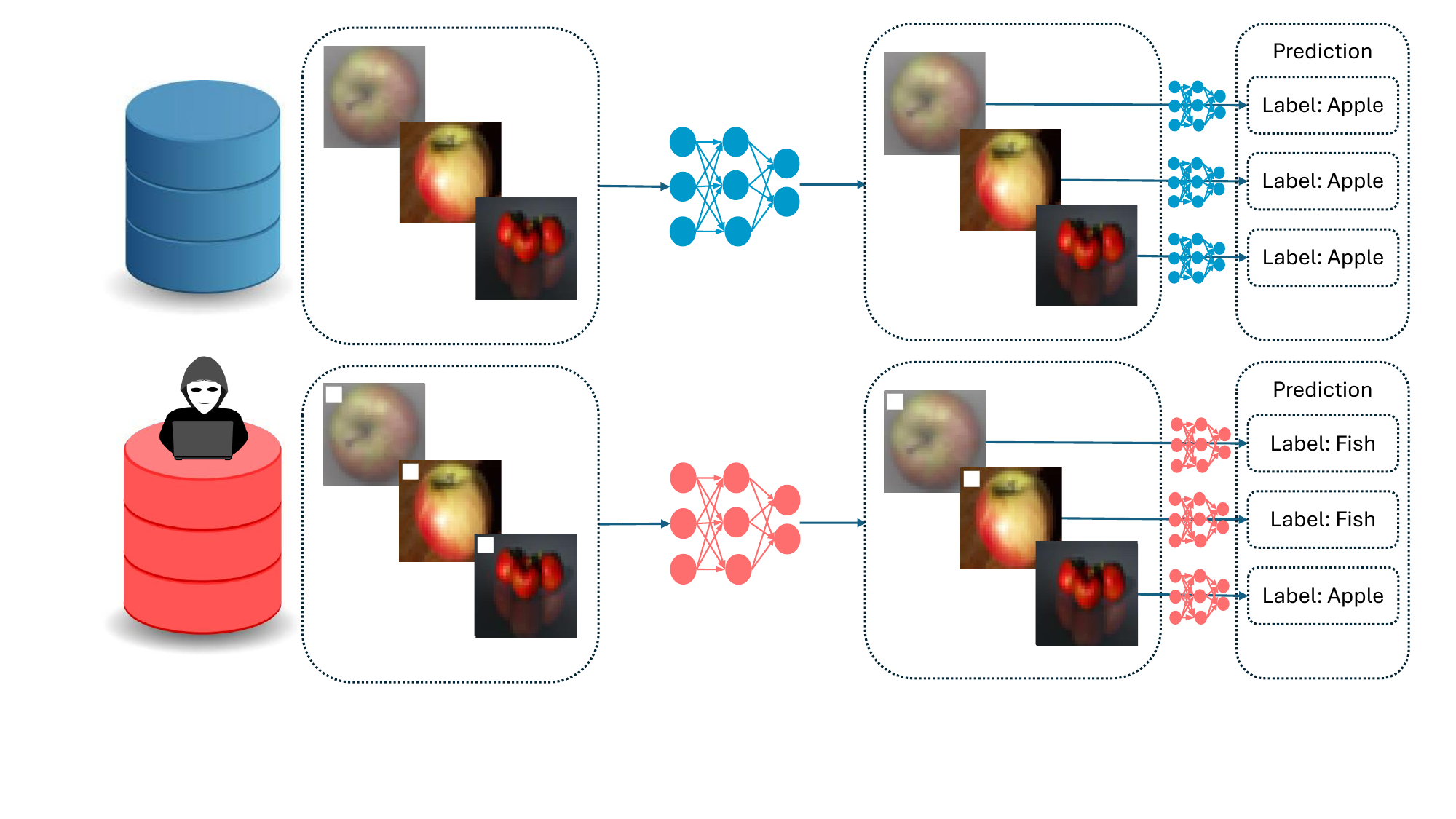}
    \caption{Figure outlining a model trained an clean data and the resultant predictions, and a model trained on targeted backdoor data, along with the resultant predictions. In this figure we have a small white square as the backdoor and when the trained model comes across this pixel formation, it predicts the target class fish.}
    \label{fig:backdoor}
\end{figure}

\subsection{Membership Inference Attacks}
Shokri et al. \cite{shokri_membership_2017} demonstrated that a malicious actor can train an attack model to infer whether specific data samples were included in the training set of a target model, thereby compromising its privacy. The key insight underlying Membership Inference Attacks (MIAs) is that machine learning models tend to respond differently to samples seen during training compared to unseen samples. Formally, for a target model $G^\tau$ that produces an output vector:
\begin{equation}
    z = G^\tau(x) \in \mathbb{R}^k
\end{equation}

To perform a membership inference attack, an adversary typically follows these steps:
\textbf{(i)} An attacker creates $R$ shadow data-sets ($D_r$) with similar distributions to the data-set used to train $G^\tau$. The data-set $D_r$ is split into \textit{in} and \textit{out} samples and the attacker knows the nature of each sample:
    \begin{equation}
        m(x) = 
        \begin{cases}
            1 \text{ if x is \textit{in}} \\
            0 \text{ if x is \textit{out}}
        \end{cases}
    \end{equation}

\textbf{(ii)} Shadow models $G^s_r$ are trained on each data-set. The attacker stores $z^S_r = G^S_r(x)$, which is the output vector across a given sample, and the nature of the sample ($m(x)$).

\textbf{(iii)} An attack model $G^A$ is then trained, with the data-set being as follows:
    \begin{equation}
        D^A = \cup^R_{r=1}\{z^S_r(x),m(x) :x \in D_r\}
    \end{equation}
    The output of this model being $G^A:\mathbb{R}^k \to \{0,1\}$

\textbf{(iv)} The attacker then queries $G^\tau$ with a given sample and uses $G^A$ to classify the output as coming from a model that was/was not trained on that sample. For a sample $x$, we have:
    \begin{equation}
        \tilde{m} = G^A(G^\tau(x))
    \end{equation}
    where $\tilde{m}$ is the predicted membership bit of the sample.

These attacks were extended upon in further papers \cite{carlini_membership_2022, zarifzadeh_low-cost_2024, ye_enhanced_2022} to be more effective, however they all consider the performance of models over samples that have been included in the training dataset and those that have not been used to train their attack model.

\section{Threat Model}
\label{OurAttack}

We consider a standard FL setting, where a central server coordinates the training process across a population of $N$ clients. In each training round, a subset of clients is randomly selected to compute local updates on their private datasets. These local updates are then sent to the server, which applies an aggregation rule (e.g., FedAvg or a RA variant) to produce the updated global model.

\noindent\textbf{Adversary Capabilities. }
We assume a subset of the clients, up to a fraction $M$, are controlled by an adversary. These adversarial clients participate in training as normal but can arbitrarily manipulate their local objectives, training procedures, or model updates before submitting them to the server. The adversary is \emph{adaptive}: it observes the global model, which is sent from the server to the clients, across rounds and updates its strategy dynamically based on self-estimated feedback. However, the adversary has \textbf{no white-box access} to the server-side aggregation rule or other benign clients’ updates. Its only observable feedback is the global model published by the server after aggregation.

\noindent\textbf{Adversary Goals. }
The adversary’s objective is to \emph{poison} the global model by injecting malicious updates that force the model to misclassify specific inputs (e.g., inputs with a trigger pattern) into a chosen target class, while maintaining high accuracy on clean data. The adversary seeks to maximize the effectiveness of the attack while remaining \emph{evasive}---that is, ensuring that malicious updates are statistically consistent with benign updates so as to avoid detection or exclusion by RA methods.


\noindent\textbf{Assumptions on Defenses. }
We assume the server employs state-of-the-art RA methods. 


\section{\ourapproach: Proposed Approach} \label{attackmethodology}

In this section, we propose \ourapproach, an adaptive poisoning framework that makes malicious client updates statistically indistinguishable from benign ones while still steering the global model in a malicious direction. The core idea is a closed-loop attack: after each round, the adversary infers —via a lightweight side-channel signal from the newly aggregated model— whether its prior contribution was incorporated, and uses this feedback to tune a local objective that balances a malicious term with a camouflage (conformity) term, keeping updates inside the aggregator’s acceptance region.

\ourapproach operates in the parameter/update space, analogously to imperceptible input-space attacks~\cite{ijcai2022p242}, but without white-box server access. The approach is aggregation-agnostic, requiring only the released global model across rounds, and can be instantiated with different proximity metrics (e.g., $\ell_2$, cosine, Huber\cite{huber_robust_1964}) for targeted backdoor goals. 

An overview of our approach is depicted in Figure~\ref{fig:FullSystem}. In each round $t$, the server broadcasts the current global model $G^{t-1}$ to a subset of clients; benign clients train locally as usual, while the adversary proceeds as follows: (i) it probes the newly published $G^{t-1}$ on a small backdoor set to obtain a side-channel signal $v_t$ estimating whether its previous contribution was incorporated; (ii) it uses this signal to a balance coefficient $\alpha_t = f(v_t)$ that controls the strength of camouflage weight; (iii) it optimizes a composite objective representing two optimisation directions, i.e., the malicious backdoor and the camouflage objective; and (iv) it submits the resulting in-distribution update to the server. The server aggregates all received updates into the new global model $G^{t}$, closing the feedback loop. In the next round, the attacker continues the optimization by deriving the new signals from the new aggregated model. 


\begin{figure}[tp]
    \centering
    \includegraphics[width=\linewidth]{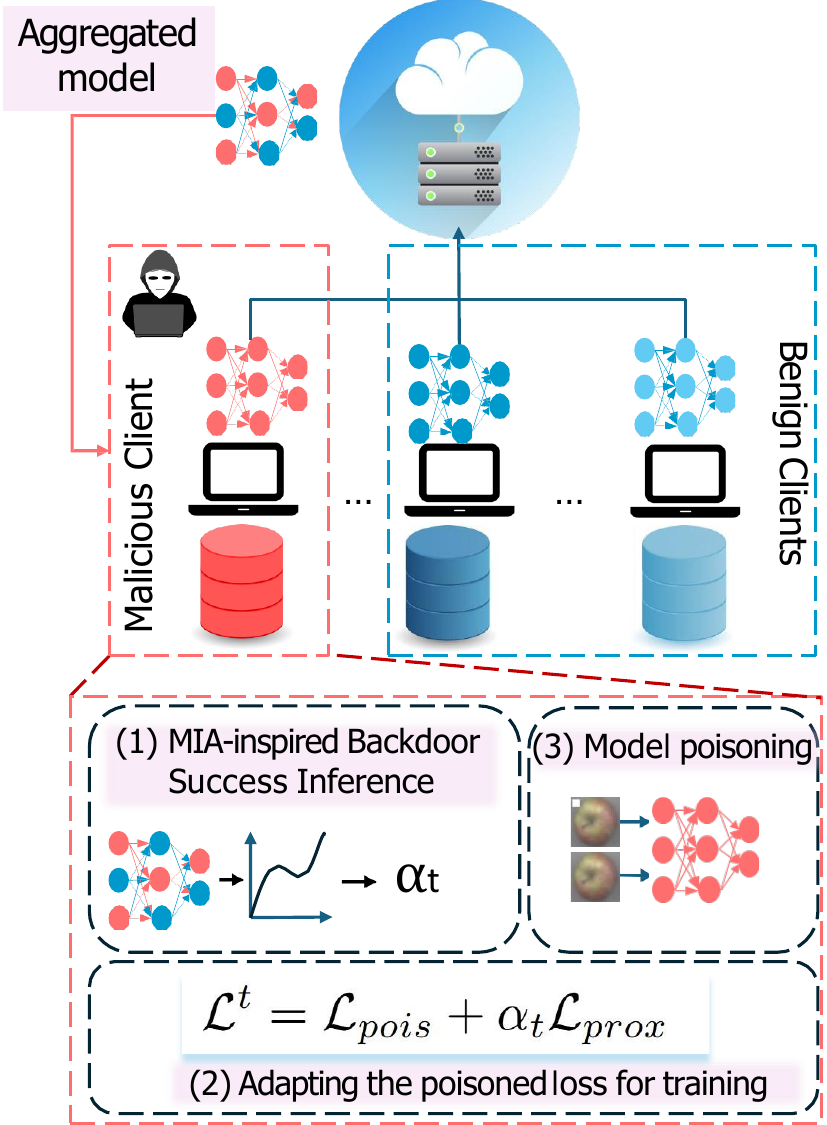}
    \caption{Outline of our \ourapproach attack with one malicious client. It depicts 3 steps: (1) the backdoor side channel inference to estimate the membership of a backdoor attack in the global model, (2) Adapting the poisoning with the malicious and the camouflage components, (3) implement model poisoning in local model.}  
    \label{fig:FullSystem}
\end{figure}

Backdoor poisoning induces a dependence between a trigger pattern and a target label by altering the training distribution. When poisoned update is incorporated within the global model, the learned predictor $F$ undergoes a behavioral shift: the conditional distribution of outputs $F(x)$ on triggered inputs differs from that on their clean counterparts (equivalently, the joint distribution over $(x,y)$ is modified within the trigger subpopulation). Inspired by the MIAs \cite{shokri_membership_2017}, this predictable shift provides a reliable probe: applying a lightweight, MIA-style test to a small set of triggered examples yields a side-channel signal indicating whether the backdoor update vector has been incorporated into the current global model. A full exploratory analysis is provided in Appendix~\ref{APP:A} to empirically validate this intuition.

The  adversarial client extracts this round-wise side-channel from $G^{t-1}$ to estimate whether its previous malicious update was incorporated or suppressed by aggregation. It then maps this estimate to a balance coefficient that adapts the local objective, increasing camouflage when incorporation appears unlikely and increasing attack strength when incorporation appears likely. This feedback loop drives the attacker to introduce as much poisoning as possible while remaining in-distribution and thus evasive to robust aggregation.



\subsection{Backdoor Side-Channel} 

Conventionally, malicious clients introduce the malicious payload in their update in hope this will be eventually added to the global model. This, however, is done without any feedback on their success. In our approach, at every iteration the adversary will estimate the success of introducing the malicious payload into the global model by estimating the membership of the poisoned samples in the global model $G^{t-1}$. 
To do this, a malicious actor performs the following operations during training:
\begin{enumerate}
    \item \textbf{Splitting the data}: At training round $t$, the malicious actor(s) gathers all of the data available to it  ($D_1,...D_M$), and splits it into datasets $d_1,...d_R$ that will be used to train $R$ reference models.
    \item \textbf{Poisoning the data-sets}: Each reference data-set is backdoored to varying degrees indicated by the percentage of source class samples $\hat{x}$ we poison. So for each reference model we have a backdooring percentage, $p_1,...,p_R$.
    \item \textbf{Training reference models}: These data-sets are used to train local reference models $h_{t,1},...,h_{t,R}$. 
    \item \textbf{Training the Backdoor Side Channel Inference (BSCI) model}: The backdoor side channel model is trained on the output vectors of the reference models across poisoned samples $z(\hat{x})$. We run each model $h_{t,r}$ for $r \in 1,...,R$ across all source class samples with the backdoor included and we label all vectors coming from models with a $p$ value of 0 as out-samples, and all other vectors as in-samples. We then train a simple SVM binary classifier as $G^t_A$ to detect vectors coming from poisoned models and those coming from non-poisoned models.
    \item \textbf{Using the model}: With the BSCI model trained, we use backdoored samples as inputs to $G^{t-1}$ and apply the BSCI to the output feature vector of $G^t$ across poisoned samples. In this manner, $G^t_A(G^{t-1}(\hat{x}))$ will return the predicted membership bit $\tilde{m}(\hat{x})$ for a backdoored sample $\hat{x}$. The average membership bit over a set of backdoored samples makes up our membership prediction ($v_t = \frac{1}{|(\hat{D}_1,\dots,\hat{D}_M)|}\sum\tilde{m}(\hat{x}$) for the published global model $G^{t-1}$.
\end{enumerate}
This Backdoor Side-Channel attack is outlined as an algorithm in Algorithm. \ref{ALG:MIE}.

\begin{algorithm*}[ht]
\caption{Backdoor Side-Channel Inference}\label{ALG:MIE}
\begin{algorithmic}[1]
 \renewcommand{\algorithmicrequire}{\textbf{Input:}}
 \renewcommand{\algorithmicensure}{\textbf{Output:}}
 \REQUIRE Data-sets $(D_1,\dots,D_M)$, $G^t$.
 \ENSURE  Estimation on likelihood of $G^t$ being backdoored.
    \STATE Split $[D_1,\dots,D_M]$ into $d_1,...d_R$.
    \STATE Poison $d_1,...d_R$ to differing percentages defined by $p_1,...,p_R$. Poisoned sample sets are noted as $\hat{D}_m \subset D_m$
    \FOR{$d_r \in d_1,...d_R$}
        \STATE Train reference models $h_{t,r}$
    \ENDFOR
    \FOR{$h_{t,r} \in h_{t,1},...h_{t,R}$}
        \STATE  Query $h_{t,r}$ using backdoored samples $\hat{x}$ and, for each sample, store the output vector $z$ and $m(\hat{x})$ the membership bit.
    \ENDFOR
    \STATE Create a dataset $D^A = \cup^R_{r=1}\{z(\hat{x}),m(\hat{x}) :\hat{x} \in (\hat{D}_1,\dots,\hat{D}_M)\}$
    \STATE Train Attack model $G^t_A$ on $D^A$
    \STATE Query $G^t$ using backdoored samples ($[\hat{D}_1,\dots,\hat{D}_M]$) and save $G^t(\hat{x})$, the output vector of the model.
    \STATE Use $G^t_A$ to obtain $\tilde{m}(\hat{x}) = G^t_A(G^t(\hat{x}))$, the membership bit prediction for each sample.
    \STATE Calculate the likelihood that $G^t$ is backdoored using $v_t = \frac{1}{|(\hat{D}_1,\dots,\hat{D}_M)|}\sum\tilde{m}(\hat{x})$.
    \RETURN $v_t$
    \end{algorithmic}
\end{algorithm*}

\subsection{Balancing loss for in-distribution attacks}

The output of the inference model for $G^t$ is a value $0\leq v_t \leq 1$ that represents an estimation of how likely a global model is to be backdoored, and is estimated by analysing it's performance across backdoored samples. The adversary can then use this aggregated estimation to update the loss function of our malicious client(s) to be more in line with the distribution of benign updates, if the previous estimation is low, or to allow for more poisoning payload, if the previous estimation value is high and therefore we are already in-distribution.

We wish to lower the proximal term if our malicious updates are impacting the global model. Therefore in our approach, we define the term $\alpha_t$ as one minus the average backdoor side channel success over the last $k$ training rounds. So for training round $t$ we have:
\begin{equation}
         \alpha_t = 1 - (\sum_{i = t - k}^t\frac{v_i}{k})
    \label{eq:alpha}
\end{equation}

Thus, the loss function of a local malicious model is altered by taking into consideration the effectiveness of the previous attack as highlighted. The loss function is adapted in relation to the following equation, for training round $t$ 
\begin{equation}
    \mathcal{L}^{t} = \mathcal{L}_{pois} + \alpha_t \mathcal{L}_{prox} 
    \label{eq:loss}
\end{equation}
where $\mathcal{L}_{poiss}$ is the classical misclassification loss over poisoned data, $\mathcal{L}_{prox}$ is the proximity to the previous rounds global model and $\alpha_t$ is the parameter that balances the proximal term at round $t$. The term $\mathcal{L}_{prox}$ is a general term for any distance metric between the malicious actors update and the average distribution of updates across all clients. In Section.\ref{Experimental Results} we investigate three specific metrics for comparison, however any metric is valid.

This works as if we have a poisonous model update that is fully included in the aggregation, our average inference success will theoretically be $1$ over $k$ rounds (therefore $\alpha=0$) and the need to alter our loss function is not required.


\section{Experimental Setup}\label{sec:exp_setup}
We evaluate our novel attack for 
targeted 
backdoor attacks against the state-of-art RA schemes highlighted in Section.~\ref{Related Work}. Experiments are ran with ten clients, one of which is malicious (10\% or $M$=1). All clients are included in every round of training, and the malicious client attacks the global model during all rounds of training. When the experiment includes a malicious actor implementing our BSCI, the following parameters are used for the adaptive attack; $R$ = 6, $p$ = \{$0.3,0.2,0.1,0.0,0.0,0.0$\}. The SVM which will be used to classify samples as members uses a polynomial kernel and a tolerance value of 0.001. Unless stated otherwise, a backdoor trigger is introduced in the top left corner of the image of size 3x3 with all white pixels. Trigger is introduced in all samples of class $y^s$ in the malicious client.

\subsection{Datasets}
Our evaluation is performed across two common image classification datasets and under IID settings:
\begin{enumerate}
    \item \textit{Cifar-10 \cite{noauthor_cifar-10_nodate}:} In this setting we simulate a FL system that trains for $T=100$ training rounds with 5 rounds of local training between each upload to the central server. We use the Alexnet model \cite{krizhevsky_imagenet_2012} in each client across the system, trained using a learning rate $\eta = 0.01$, and a batch size $B = 64$. We use the parameter $k$ = 5 for our adaptive attack in this setting. The data is not augmented, except for samples which are backdoored.
    \item \textit{Fashion-MNIST \cite{xiao_fashion-mnist_2017}:} In this setting train for $T=50$ global training rounds with 5 local training rounds. We used a higher learning rate of $\eta = 0.1$, and used a batch size $B = 64$. We use a lightweight CNN modified specifically for the Fashion-MNIST dataset, this is outlined in \ref{APP:C}. In this setting, we use $k = 3$, and once again, the data is not augmented except for backdooring source class samples in the malicious client(s).
\end{enumerate}

\subsection{Metrics}
To measure the success of the targeted backdoor attacks, we focus our analysing on the attack success rate (ASR). We calculate ASR for a model $G$ trained on a data-set $D$ with a set backdoored samples $\hat{X}$ aiming to misclassify source class $y^s$ to $y^\tau$ as:
\begin{equation}
    \text{ASR} = \frac{\#G(\hat{X}) = y^\tau}{\# \hat{X}}
\end{equation}
As reference, we also review the accuracy of the global model across benign clients, as backdoors should have minimal impact over benign samples so that the average accuracy should remain comparable to when there is no attack. Some RA defences also return values which make it easier for us to review how well our attack performs in terms of getting into the distribution of the aggregator, such as Krum selecting one benign/malicious client at every training round. When this is available, we also analyse these values.

\subsection{RA Scheme Parameters}
To facilitate the best possible defence, we setup each defence to be theoretically robust against one client in the system. Note that FedAvg, Median and FoolsGold have no parameters to be altered. The parameters of each defence used in our experiments are as follows:
\begin{itemize}
    \item[] Trimmed-Mean: $\beta = 0.2$
    \item[] Krum: $f =1$
    \item[] Multi-Krum: $m=3$,$f =1$
    \item[] Bulyan: $m=3$,$f=1$
    \item[] RFA: $10$ iterations max, $\epsilon = 1e^{-10}, tol=1e^{-5}$
    \item[] AlignIns/DAI: $th=0.1$
    \item[] RLR: $c=1$,$\theta=1$
\end{itemize}


\section{Experimental Results}\label{Experimental Results}
In this section, we investigate the performance of our chameleon adaptive poisoning attack \ourapproach. First, the efficacy of approach is evaluated against state-of-art RA defences, and compared with a baseline consisting in a classical vanilla data poisoning attack. Then, a set of ablation experiments and detailed analysis are carried out to investigate the most relevant factors affecting the performance of our novel attack, such as the metric distance used to calculate the attack updates within the acceptance region, the size of the backdoor trigger and the effectiveness of our BSCI mechanism in comparison with the ASR.

\subsection{Attack efficacy and baseline comparison}
In this experiment, \ourapproach is benchmarked against a vanilla backdoor attack using the parameters and settings by default described in Section.\ref{sec:exp_setup}. For \ourapproach, the best proximity metric is used in every case among three options -Euclidean distance, cosine similarity and Huber loss-. We do not launch our informed attack against FedAvg scheme, as all clients are already inside the distribution by definition (all clients are included in aggregation) so there is no need for a proximity metric in the loss function.

Table.\ref{table:overall} summarises the results for a 3x3 pixel backdoor trigger in both datasets and against all RA methods outlined in Section.\ref{Related Work}, wiht respect to ASR. In all reported cases, \ourapproach outperforms the baseline, except on the cases of Median, FTA and RFA in FashionMNIST, where the performance is almost equivalent. In 5 out of the 9 RA schemes, our approach is fully successful on evading the RA. In those cases, the poisoning against RA defenses even outperformed a poisoning attack against no defence (FedAvg). RA schemes Krum, Multi-Krum, Bulyan and DAI, which perform best against the baseline compared to the other schemes, also show our biggest gains.
The set of RA schemes that use differing methods (see Section.\ref{Related Work}) to create a set of acceptable clients for aggregation. RA schemes of this nature seem most susceptible to our attack, as we see on average an increase in ASR across these settings of $69.1981\times$. The RA scheme RLR does not select clients in a similar fashion to the previous schemes, however, our stealthy approach is able to circumvent this defence also.

Median and Trimmed-Mean are both aggregation schemes that calculate the new global model parameter-wise. Because of this it is harder to estimate the affect that a malicious client has on the global model so that our inference model cannot make strong predictions on the membership of a backdoor in the global model. However, they are not as robust as other defences in this setting against the baseline attack.

Only on those cases where the defence was unsuccessful ($ASR > 24\%$) in the first place at stopping the vanilla attack (e.g. RFA, Foolsgold) does our method not provide a significant increase. This is due to the malicious updates to be already in-distribution of these weak defences. 

Average accuracy on the benign clients (Benign-ACC) is also reported in this table to illustrate how our attack does not impact the overall accuracy beyond the backdoored samples.

\begin{table*}[h!]
\centering
\begin{tabular}{c|cccc|cccc|}
\cline{2-9}
 &
  \multicolumn{4}{c|}{Fashion-MNIST} &
  \multicolumn{4}{c|}{Cifar-10} \\ \cline{2-9} 
 &
  \multicolumn{2}{c|}{Data-Poisoning} &
  \multicolumn{2}{c|}{\ourapproach} &
  \multicolumn{2}{c|}{Data-Poisoning} &
  \multicolumn{2}{c|}{\ourapproach} \\ \hline
\multicolumn{1}{|c|}{Defence} &
  \multicolumn{1}{c|}{Benign-ACC} &
  \multicolumn{1}{c|}{ASR} &
  \multicolumn{1}{c|}{Benign-ACC} &
  ASR &
  \multicolumn{1}{c|}{Benign-ACC} &
  \multicolumn{1}{c|}{ASR} &
  \multicolumn{1}{c|}{Benign-ACC} &
  ASR \\ \hline \hline
\multicolumn{1}{|c|}{FedAvg} &
  \multicolumn{1}{c|}{91.7000} &
  \multicolumn{1}{c|}{99.1820} &
  \cellcolor[HTML]{C0C0C0}{\color[HTML]{C0C0C0} } &
  \cellcolor[HTML]{C0C0C0}{\color[HTML]{C0C0C0} } &
  \multicolumn{1}{c|}{62.1000} &
  \multicolumn{1}{c|}{50.5660} &
  \cellcolor[HTML]{C0C0C0}{\color[HTML]{9B9B9B} } &
  \cellcolor[HTML]{C0C0C0}{\color[HTML]{9B9B9B} } \\ \hline
\multicolumn{1}{|c|}{Median} &
  \multicolumn{1}{c|}{91.8000} &
  \multicolumn{1}{c|}{\textbf{24.3354}} &
  \multicolumn{1}{c|}{91.8000} &
  20.4450 &
  \multicolumn{1}{c|}{57.4000} &
  \multicolumn{1}{c|}{18.4049} &
  \multicolumn{1}{c|}{57.1000} &
  \textbf{29.4340} \\ \hline
\multicolumn{1}{|c|}{FTA} &
  \multicolumn{1}{c|}{91.9000} &
  \multicolumn{1}{c|}{\textbf{28.4254}} &
  \multicolumn{1}{c|}{91.7000} &
  26.1759 &
  \multicolumn{1}{c|}{61.7000} &
  \multicolumn{1}{c|}{23.1084} &
  \multicolumn{1}{c|}{61.3000} &
  \textbf{37.3595} \\ \hline
\multicolumn{1}{|c|}{Krum} &
  \multicolumn{1}{c|}{88.2000} &
  \multicolumn{1}{c|}{0.2045} &
  \multicolumn{1}{c|}{81.7000} &
  \textbf{99.7955} &
  \multicolumn{1}{c|}{47.9000} &
  \multicolumn{1}{c|}{3.0675} &
  \multicolumn{1}{c|}{47.7000} &
  \textbf{100} \\ \hline
\multicolumn{1}{|c|}{MKrum} &
  \multicolumn{1}{c|}{89.3000} &
  \multicolumn{1}{c|}{0.4090} &
  \multicolumn{1}{c|}{88.2000} &
  \textbf{99.1820} &
  \multicolumn{1}{c|}{57.3000} &
  \multicolumn{1}{c|}{1.4315} &
  \multicolumn{1}{c|}{56.4000} &
  \textbf{83.0189} \\ \hline
\multicolumn{1}{|c|}{Bulyan} &
  \multicolumn{1}{c|}{89.7000} &
  \multicolumn{1}{c|}{0.0000} &
  \multicolumn{1}{c|}{89.4000} &
  \textbf{96.1145} &
  \multicolumn{1}{c|}{60.0000} &
  \multicolumn{1}{c|}{0.8180} &
  \multicolumn{1}{c|}{57.1000} &
  \textbf{82.0755} \\ \hline
\multicolumn{1}{|c|}{DAI} &
  \multicolumn{1}{c|}{91.4000} &
  \multicolumn{1}{c|}{0.2045} &
  \multicolumn{1}{c|}{90.6000} &
  \textbf{98.3640} &
  \multicolumn{1}{c|}{60.0000} &
  \multicolumn{1}{c|}{2.4528} &
  \multicolumn{1}{c|}{60.5000} &
  \textbf{33.5849} \\ \hline
\multicolumn{1}{|c|}{RLR} &
  \multicolumn{1}{c|}{92.3000} &
  \multicolumn{1}{c|}{29.8569} &
  \multicolumn{1}{c|}{87.7000} &
  \textbf{99.7955} &
  \multicolumn{1}{c|}{58.7000} &
  \multicolumn{1}{c|}{26.2264} &
  \multicolumn{1}{c|}{53.9000} &
  \textbf{97.3585} \\ \hline
\multicolumn{1}{|c|}{RFA} &
  \multicolumn{1}{c|}{91.8000} &
  \multicolumn{1}{c|}{\textbf{99.7955}} &
  \multicolumn{1}{c|}{91.6000} &
  97.9550 &
  \multicolumn{1}{c|}{61.0000} &
  \multicolumn{1}{c|}{43.3962} &
  \multicolumn{1}{c|}{60.6000} &
  \textbf{45.6604} \\ \hline
\multicolumn{1}{|c|}{FoolsGold} &
  \multicolumn{1}{c|}{91.3000} &
  \multicolumn{1}{c|}{\textbf{98.7730}} &
  \multicolumn{1}{c|}{91.0000} &
   98.5685 &
  \multicolumn{1}{c|}{60.8000} &
  \multicolumn{1}{l|}{46.9811} &
  \multicolumn{1}{c|}{62.0000} &
  \multicolumn{1}{l|}{\textbf{47.5472}} \\ \hline
\end{tabular}
\caption{Comparative results of our proposed adaptive attack \ourapproach versus a vanilla data poisoning. In bold we represent the best ASR performance for every dataset and RA scheme.
}
\label{table:overall}
\end{table*}

\subsection{Ablation experiment: Backdoor trigger size}

Figure~\ref{fig:bestASRS} represents graphically the comparison from Table~\ref{table:overall}, extending it for different strength of backdooring, expressed as the size of the backdoored trigger. Similar conclusions can be extracted for all sizes, with our adaptive attack consistently outperforming the baseline, expect on those cases where the baseline is already effective against RA scheme.

\begin{figure*}
    \centering
    \begin{subfigure}{0.32\textwidth}
        \includegraphics[width=\linewidth]{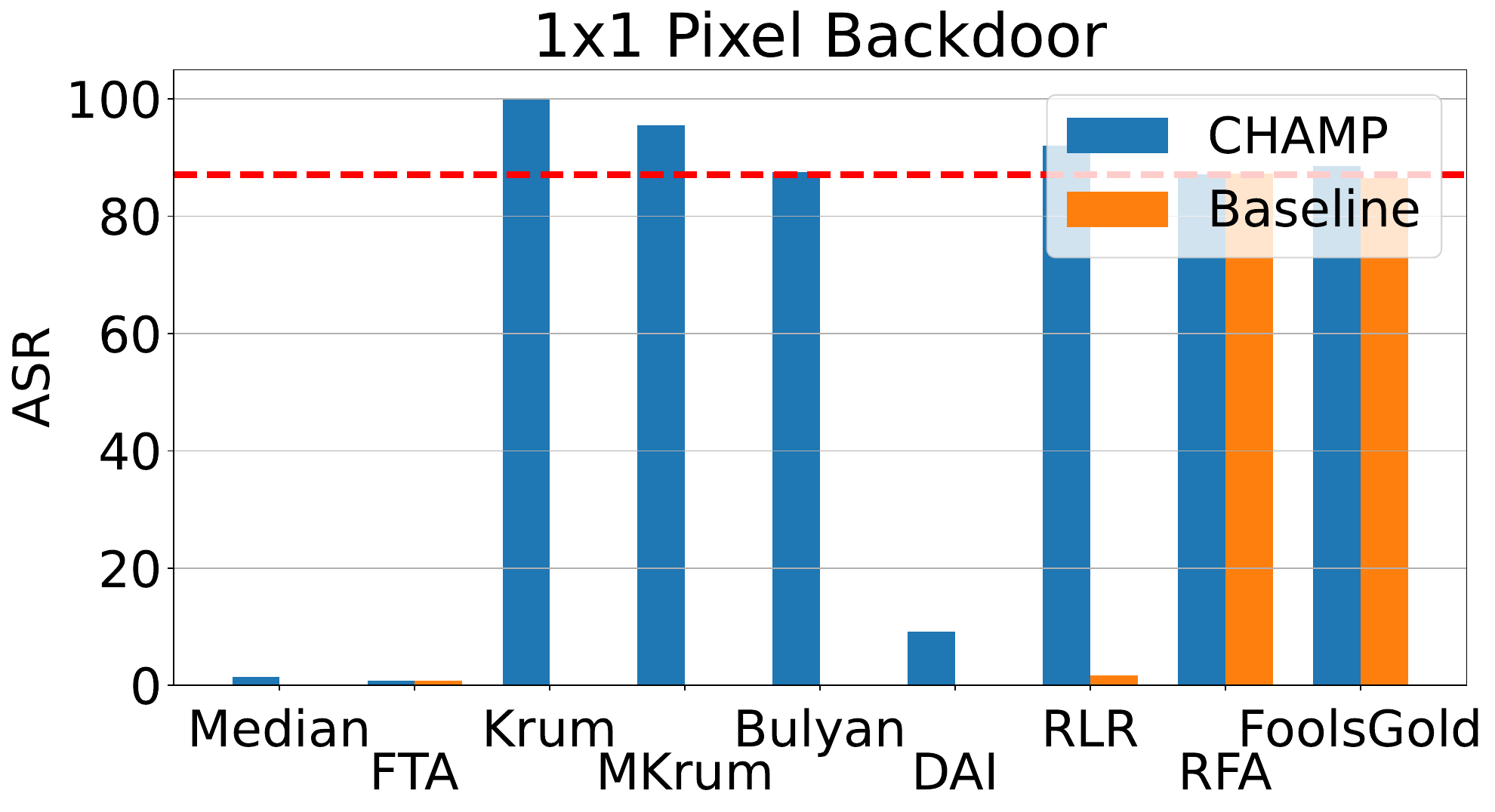}
    \end{subfigure}
    \hfill
    \begin{subfigure}{0.32\textwidth}
        \includegraphics[width=\linewidth]{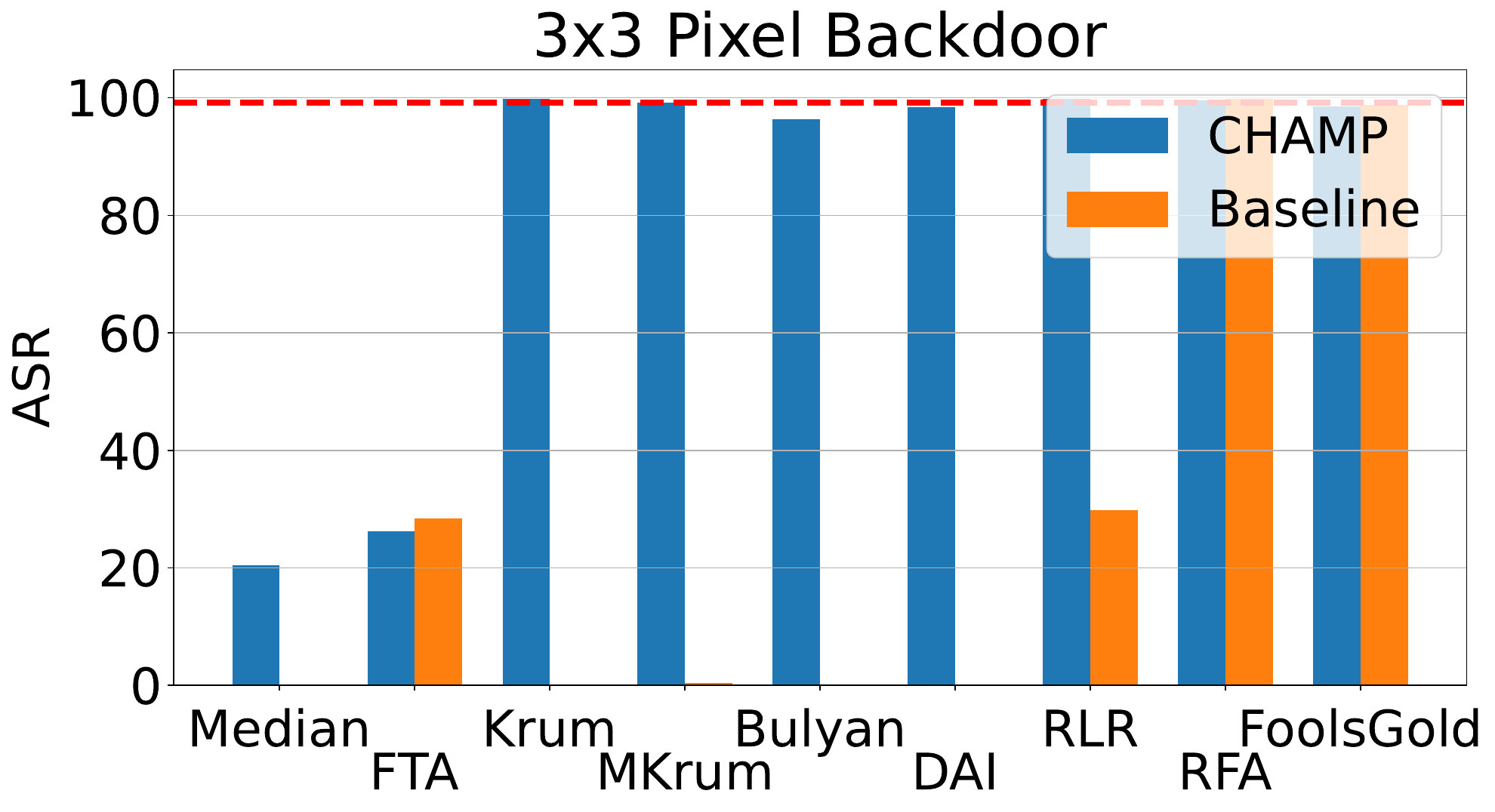}
    \end{subfigure}
    \hfill
    \begin{subfigure}{0.32\textwidth}
        \includegraphics[width=\linewidth]{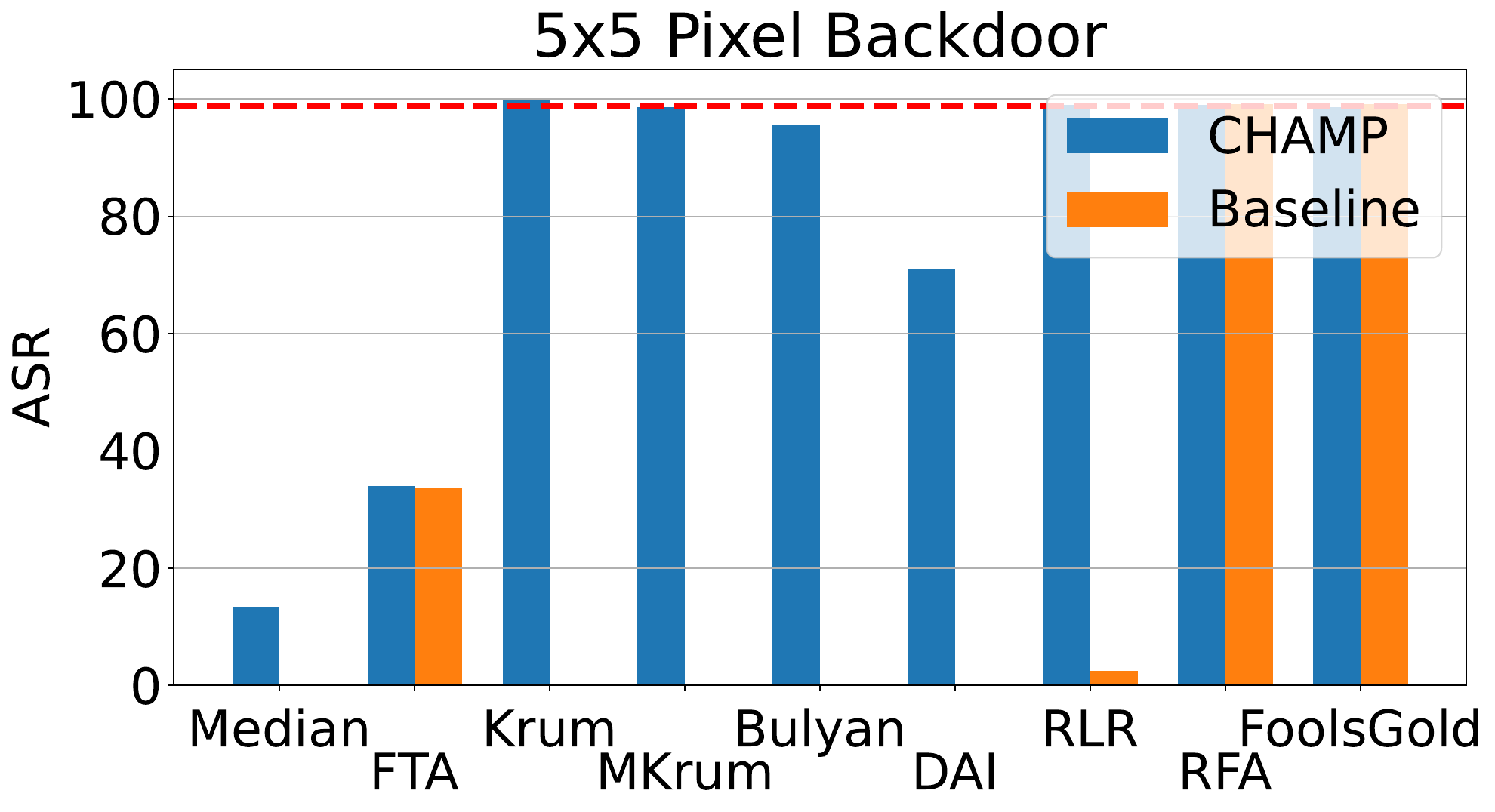}
    \end{subfigure}
    
    \vspace{0.5em}
    \begin{subfigure}{0.32\textwidth}
        \includegraphics[width=\linewidth]{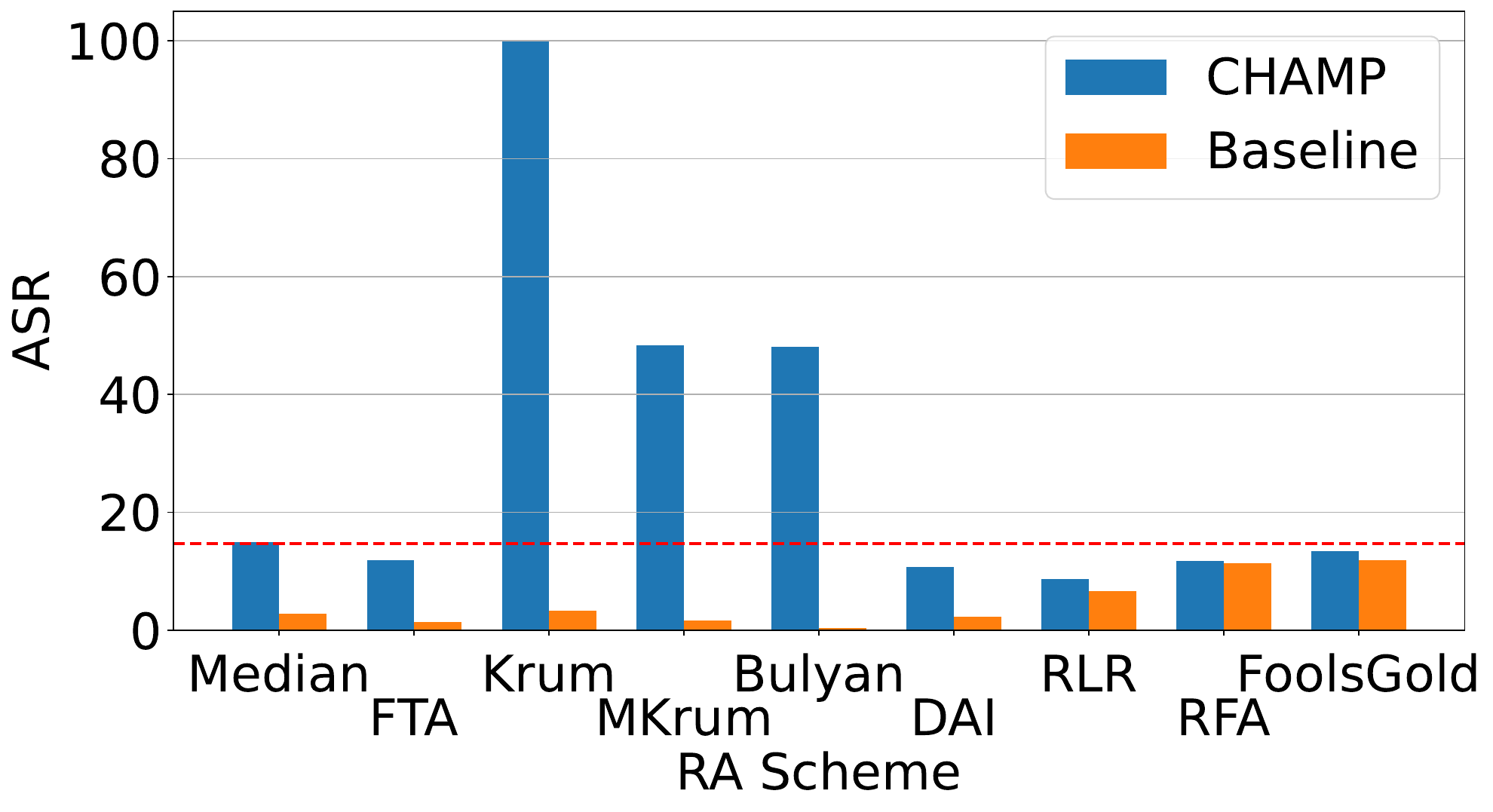}
    \end{subfigure}
    \hfill
    \begin{subfigure}{0.32\textwidth}
        \includegraphics[width=\linewidth]{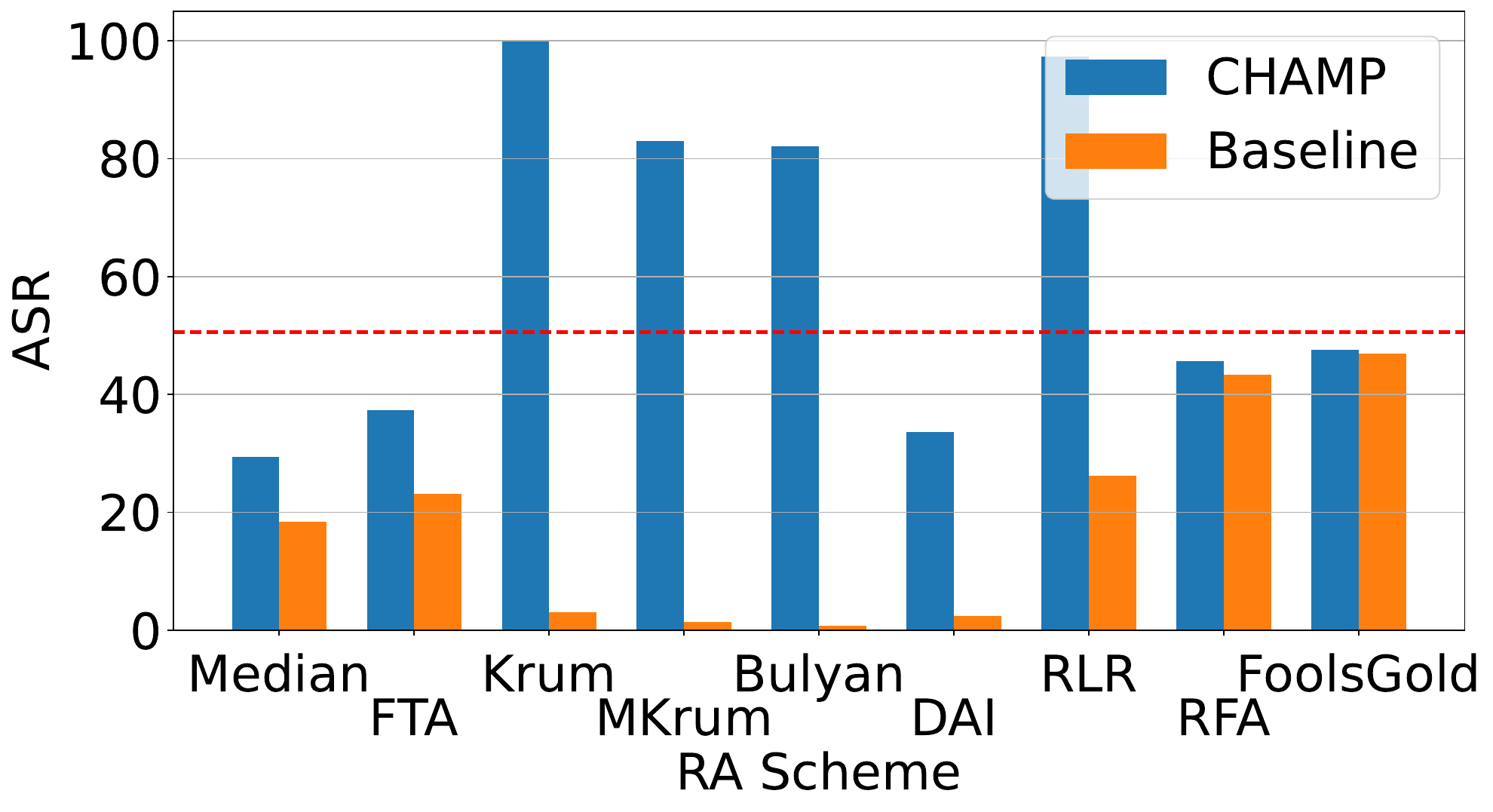}
    \end{subfigure}
    \hfill
    \begin{subfigure}{0.32\textwidth}
        \includegraphics[width=\linewidth]{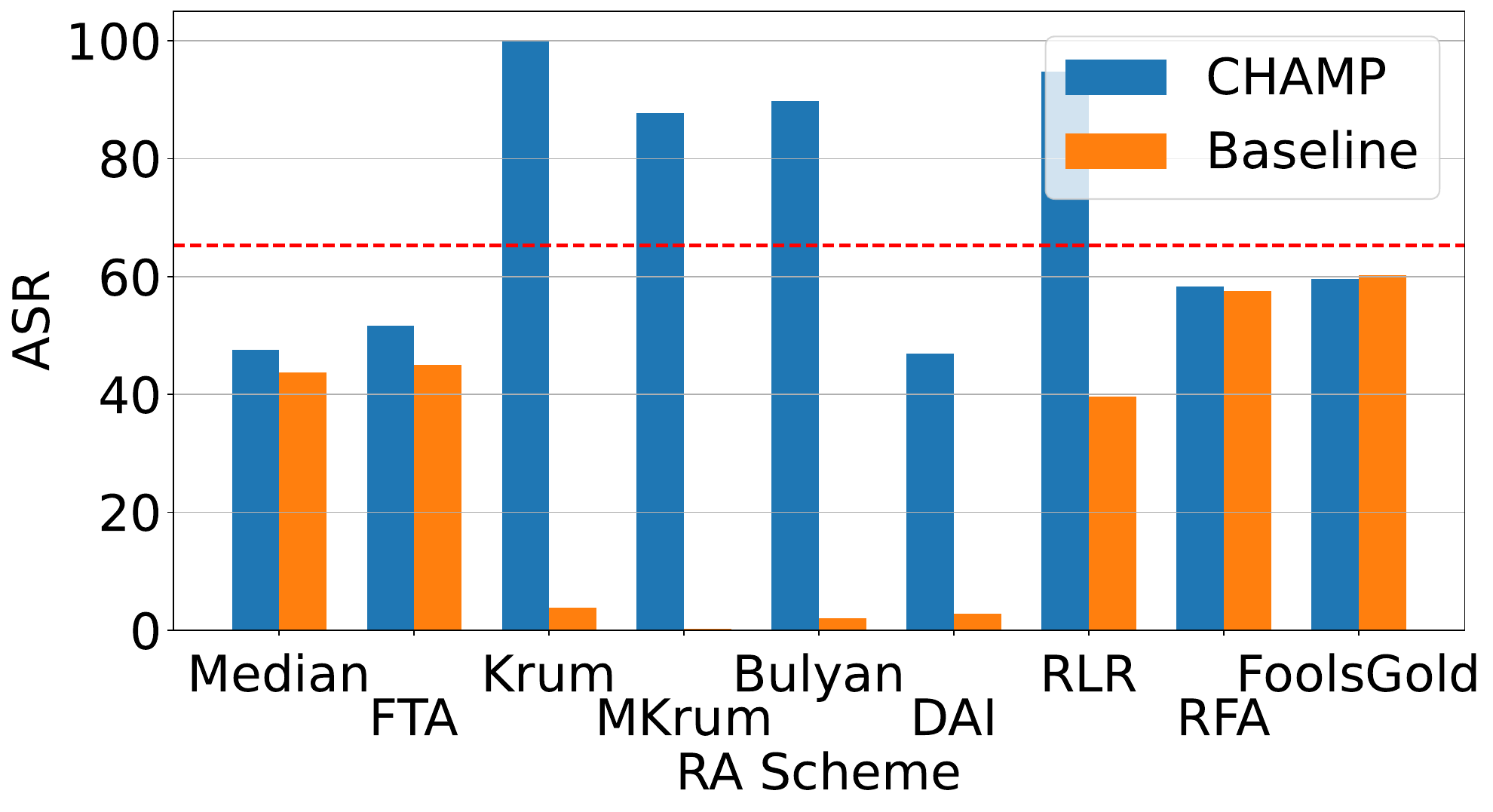}
    \end{subfigure}
    \caption{ASR comparison between our adaptive attack \ourapproach and the baseline attack. The three plots represent the results for the 1x1, 3x3, and 5x5 pixel backdoor attacks on  Fashion-MNIST (first row) and Cifar-10 (Second row). The ASR of the baseline attack against FedAvg is plotted across each plot as a dashed red line as reference.} 
    \label{fig:bestASRS}
\end{figure*}

From these results we can conclude that even state-of-art RA schemes remain vulnerable to adaptive adversaries, challenging the prevailing assumption that RA ensures resilience.


In this experiment, we depict, for different trigger sizes -1x1, 3x3 and 5x5- both the global model accuracy across benign samples and ASR across backdoored samples over the training rounds. This ablation experiment is performed on both datasets. 

As expected, we can see in Figure~\ref{fig:FedAvg} that increasing the size of the backdoor improves the ASR of the attack on a normal setting using FedAvg, whilst having little to no affect on the accuracy across benign samples. Similar conclusions regarding the strength on the attack as a function of the trigger size can be extrapolated for all defenses and our adaptive attack, as plotted in Figure~\ref{fig:bestASRS}.

\begin{figure}
    \centering
    \includegraphics[width=\linewidth]{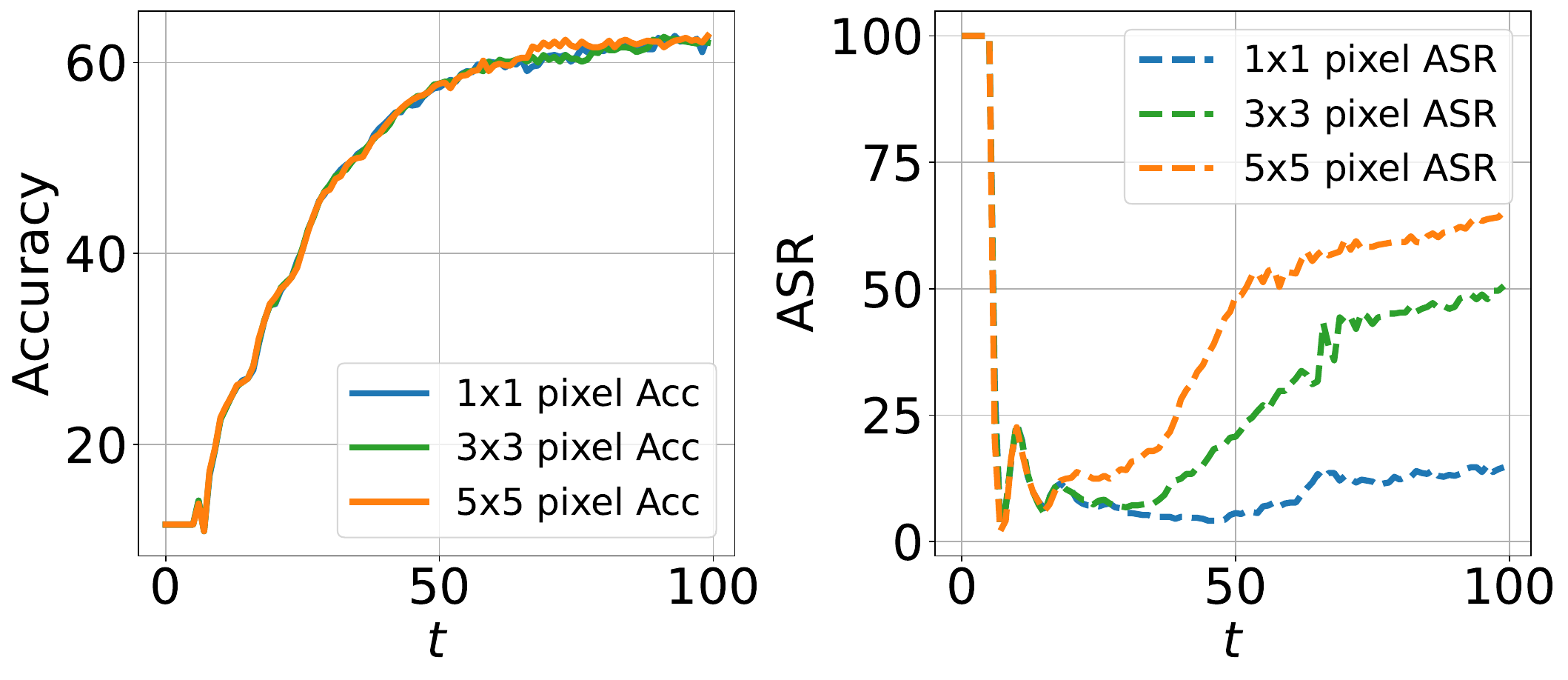}
    \caption{Global model accuracy and ASR for a vanilla backdoor attack of varying sizes against a FL system using FedAvg in the aggregation.}
    \label{fig:FedAvg}
\end{figure}

\subsection{Ablation experiment: Proximity metric}

In this experiment, we explore the influence of the proximity metric between malicious updates and the averaged update data distribution used to calculate $\mathcal{L}_{prox}$. Experiment is ocnductedon the CIFAR-10 dataset, where we compare the ASR at time $T$ and at time $\nicefrac{T}{2}$ when using a baseline data poisoning attack, compared to our adaptive attack under different proximity metrics. We use three different proximity metrics for these experiments in the attacking clients loss function, Euclidean distance, cosine similarity and the Huber loss \cite{huber_robust_1964} (all with respect to the previous rounds global model). The three backdoor trigger sizes are reported for comparison. 

Table.\ref{Tab:ScalingBackdoor} shows the results of these experiments. It can be seen how Euclidean distance and Huber loss exhibit in general a similar behavior with little differences, such as Huber being slightly better against Bulyan and DAI, while Euclidean surpasses Huber in Multi-Krum and RLR. Cosine similarity out-performs the others for specific RAs such as Median, FTA, RFA and FoolsGold, but at the cost of failing to show any improvement regarding the baseline in other five schemes. In general, in a blackbox setting where no knowledge of the aggregation rule is available, Huber loss seems the more robust approach, while efficacy gains are possible in a white-box threat model.

\begin{table*}[ht]
\begin{tabular}{|ccccccccccc|}
\hline
\multicolumn{1}{|c|}{\textbf{\begin{tabular}[c]{@{}c@{}}Pixel\\ Backdoor\end{tabular}}} &
  \multicolumn{1}{c|}{\textbf{ASR @ $t$}} &
  \multicolumn{1}{c|}{\textbf{Median}} &
  \multicolumn{1}{c|}{\textbf{FTA}} &
  \multicolumn{1}{c|}{\textbf{Krum}} &
  \multicolumn{1}{c|}{\textbf{Multi-Krum}} &
  \multicolumn{1}{c|}{\textbf{Bulyan}} &
  \multicolumn{1}{c|}{\textbf{DAI}} &
  \multicolumn{1}{c|}{\textbf{RFA}} &
  \multicolumn{1}{c|}{\textbf{RLR}} &
  \textbf{FoolsGold} \\ \hline
\multicolumn{11}{|c|}{\textbf{Euclidean Distance}} \\ \hline
\multicolumn{1}{|c|}{} &
  \multicolumn{1}{c|}{$t=\nicefrac{T}{2}$} &
  \multicolumn{1}{c|}{3.2720} &
  \multicolumn{1}{c|}{3.8855} &
  \multicolumn{1}{c|}{95.8491} &
  \multicolumn{1}{c|}{6.9811} &
  \multicolumn{1}{c|}{4.1509} &
  \multicolumn{1}{c|}{2.4528} &
  \multicolumn{1}{c|}{2.8302} &
  \multicolumn{1}{c|}{5.0943} &
  2.4528 \\ \cline{2-11} 
\multicolumn{1}{|c|}{\multirow{-2}{*}{\textbf{1x1}}} &
  \multicolumn{1}{c|}{\textbf{$t=T$}} &
  \multicolumn{1}{c|}{2.8630} &
  \multicolumn{1}{c|}{1.4315} &
  \multicolumn{1}{c|}{\textbf{100.0000}} &
  \multicolumn{1}{c|}{\textbf{63.2075}} &
  \multicolumn{1}{c|}{43.2075} &
  \multicolumn{1}{c|}{\textbf{7.7358}} &
  \multicolumn{1}{c|}{\cellcolor[HTML]{FFFFFF}\textbf{11.6981}} &
  \multicolumn{1}{c|}{\cellcolor[HTML]{FFFFFF}8.1132} &
  \cellcolor[HTML]{FFFFFF}6.2264 \\ \hline
\multicolumn{1}{|c|}{} &
  \multicolumn{1}{c|}{$t=\nicefrac{T}{2}$} &
  \multicolumn{1}{c|}{5.8491} &
  \multicolumn{1}{c|}{3.5849} &
  \multicolumn{1}{c|}{99.8113} &
  \multicolumn{1}{c|}{30.9434} &
  \multicolumn{1}{c|}{16.6038} &
  \multicolumn{1}{c|}{3.9623} &
  \multicolumn{1}{c|}{7.9245} &
  \multicolumn{1}{c|}{12.0755} &
  4.5283 \\ \cline{2-11} 
\multicolumn{1}{|c|}{\multirow{-2}{*}{\textbf{3x3}}} &
  \multicolumn{1}{c|}{\textbf{$t=T$}} &
  \multicolumn{1}{c|}{26.6038} &
  \multicolumn{1}{c|}{22.0755} &
  \multicolumn{1}{c|}{\textbf{100.0000}} &
  \multicolumn{1}{c|}{\textbf{88.8679}} &
  \multicolumn{1}{c|}{79.0566} &
  \multicolumn{1}{c|}{27.1698} &
  \multicolumn{1}{c|}{42.6415} &
  \multicolumn{1}{c|}{\textbf{97.3585}} &
  25.6604 \\ \hline
\multicolumn{1}{|c|}{} &
  \multicolumn{1}{c|}{$t=\nicefrac{T}{2}$} &
  \multicolumn{1}{c|}{16.0377} &
  \multicolumn{1}{c|}{12.2642} &
  \multicolumn{1}{c|}{99.8113} &
  \multicolumn{1}{c|}{45.2830} &
  \multicolumn{1}{c|}{34.5283} &
  \multicolumn{1}{c|}{8.1132} &
  \multicolumn{1}{c|}{23.2075} &
  \multicolumn{1}{c|}{39.8113} &
  13.3962 \\ \cline{2-11} 
\multicolumn{1}{|c|}{\multirow{-2}{*}{\textbf{5x5}}} &
  \multicolumn{1}{c|}{\textbf{$t=T$}} &
  \multicolumn{1}{c|}{44.5283} &
  \multicolumn{1}{c|}{36.2264} &
  \multicolumn{1}{c|}{\textbf{100.0000}} &
  \multicolumn{1}{c|}{\textbf{92.6415}} &
  \multicolumn{1}{c|}{86.6038} &
  \multicolumn{1}{c|}{42.2642} &
  \multicolumn{1}{c|}{56.4151} &
  \multicolumn{1}{c|}{\textbf{94.7170}} &
  37.1698 \\ \hline
\multicolumn{11}{|c|}{\textbf{Cosine Similarity}} \\ \hline
\multicolumn{1}{|c|}{} &
  \multicolumn{1}{c|}{$t=\nicefrac{T}{2}$} &
  \multicolumn{1}{c|}{2.2642} &
  \multicolumn{1}{c|}{2.2642} &
  \multicolumn{1}{c|}{2.2642} &
  \multicolumn{1}{c|}{2.2642} &
  \multicolumn{1}{c|}{2.8302} &
  \multicolumn{1}{c|}{1.8868} &
  \multicolumn{1}{c|}{3.2075} &
  \multicolumn{1}{c|}{3.9623} &
  2.6415 \\ \cline{2-11} 
\multicolumn{1}{|c|}{\multirow{-2}{*}{\textbf{1x1}}} &
  \multicolumn{1}{c|}{\textbf{$t=T$}} &
  \multicolumn{1}{c|}{\textbf{14.9057}} &
  \multicolumn{1}{c|}{\textbf{11.8868}} &
  \multicolumn{1}{c|}{2.4528} &
  \multicolumn{1}{c|}{2.0755} &
  \multicolumn{1}{c|}{1.8868} &
  \multicolumn{1}{c|}{2.6415} &
  \multicolumn{1}{c|}{\textbf{11.6981}} &
  \multicolumn{1}{c|}{\textbf{8.6792}} &
  \textbf{13.3962} \\ \hline
\multicolumn{1}{|c|}{} &
  \multicolumn{1}{c|}{$t=\nicefrac{T}{2}$} &
  \multicolumn{1}{c|}{6.4151} &
  \multicolumn{1}{c|}{4.9057} &
  \multicolumn{1}{c|}{2.2614} &
  \multicolumn{1}{c|}{2.4528} &
  \multicolumn{1}{c|}{2.6415} &
  \multicolumn{1}{c|}{1.8868} &
  \multicolumn{1}{c|}{10.5660} &
  \multicolumn{1}{c|}{10.7547} &
  12.6415 \\ \cline{2-11} 
\multicolumn{1}{|c|}{\multirow{-2}{*}{\textbf{3x3}}} &
  \multicolumn{1}{c|}{\textbf{$t=T$}} &
  \multicolumn{1}{c|}{\textbf{29.4340}} &
  \multicolumn{1}{c|}{\textbf{37.3585}} &
  \multicolumn{1}{c|}{2.4528} &
  \multicolumn{1}{c|}{1.5094} &
  \multicolumn{1}{c|}{3.5849} &
  \multicolumn{1}{c|}{2.4528} &
  \multicolumn{1}{c|}{\textbf{45.6604}} &
  \multicolumn{1}{c|}{27.3585} &
  \textbf{47.5472} \\ \hline
\multicolumn{1}{|c|}{} &
  \multicolumn{1}{c|}{$t=\nicefrac{T}{2}$} &
  \multicolumn{1}{c|}{18.1132} &
  \multicolumn{1}{c|}{15.0943} &
  \multicolumn{1}{c|}{0.9434} &
  \multicolumn{1}{c|}{2.0755} &
  \multicolumn{1}{c|}{1.6981} &
  \multicolumn{1}{c|}{2.0755} &
  \multicolumn{1}{c|}{25.4717} &
  \multicolumn{1}{c|}{25.0943} &
  33.2075 \\ \cline{2-11} 
\multicolumn{1}{|c|}{\multirow{-2}{*}{\textbf{5x5}}} &
  \multicolumn{1}{c|}{\textbf{$t=T$}} &
  \multicolumn{1}{c|}{\textbf{47.5472}} &
  \multicolumn{1}{c|}{\textbf{51.6981}} &
  \multicolumn{1}{c|}{1.5094} &
  \multicolumn{1}{c|}{2.2642} &
  \multicolumn{1}{c|}{2.0755} &
  \multicolumn{1}{c|}{3.2075} &
  \multicolumn{1}{c|}{\textbf{58.3019}} &
  \multicolumn{1}{c|}{39.2453} &
  59.6226 \\ \hline
\multicolumn{11}{|c|}{\textbf{Huber Loss}} \\ \hline
\multicolumn{1}{|c|}{} &
  \multicolumn{1}{c|}{$t=\nicefrac{T}{2}$} &
  \multicolumn{1}{c|}{3.2075} &
  \multicolumn{1}{c|}{2.4528} &
  \multicolumn{1}{c|}{94.7170} &
  \multicolumn{1}{c|}{6.0378} &
  \multicolumn{1}{c|}{4.1509} &
  \multicolumn{1}{c|}{2.4528} &
  \multicolumn{1}{c|}{2.8302} &
  \multicolumn{1}{c|}{2.8302} &
  2.6415 \\ \cline{2-11} 
\multicolumn{1}{|c|}{\multirow{-2}{*}{\textbf{1x1}}} &
  \multicolumn{1}{c|}{\textbf{$t=T$}} &
  \multicolumn{1}{c|}{12.2642} &
  \multicolumn{1}{c|}{7.3585} &
  \multicolumn{1}{c|}{\textbf{100.0000}} &
  \multicolumn{1}{c|}{48.3019} &
  \multicolumn{1}{c|}{\textbf{48.1132}} &
  \multicolumn{1}{c|}{1.5094} &
  \multicolumn{1}{c|}{10.5660} &
  \multicolumn{1}{c|}{1.8868} &
  9.2453 \\ \hline
\multicolumn{1}{|c|}{} &
  \multicolumn{1}{c|}{$t=\nicefrac{T}{2}$} &
  \multicolumn{1}{c|}{4.1509} &
  \multicolumn{1}{c|}{4.1509} &
  \multicolumn{1}{c|}{100.0000} &
  \multicolumn{1}{c|}{16.9811} &
  \multicolumn{1}{c|}{18.1132} &
  \multicolumn{1}{c|}{4.1509} &
  \multicolumn{1}{c|}{8.6792} &
  \multicolumn{1}{c|}{16.9811} &
  6.0377 \\ \cline{2-11} 
\multicolumn{1}{|c|}{\multirow{-2}{*}{\textbf{3x3}}} &
  \multicolumn{1}{c|}{\textbf{$t=T$}} &
  \multicolumn{1}{c|}{27.3585} &
  \multicolumn{1}{c|}{26.9811} &
  \multicolumn{1}{c|}{\textbf{100.0000}} &
  \multicolumn{1}{c|}{83.0189} &
  \multicolumn{1}{c|}{\textbf{82.0755}} &
  \multicolumn{1}{c|}{\textbf{33.5849}} &
  \multicolumn{1}{c|}{45.4717} &
  \multicolumn{1}{c|}{75.0943} &
  33.9623 \\ \hline
\multicolumn{1}{|c|}{} &
  \multicolumn{1}{c|}{$t=\nicefrac{T}{2}$} &
  \multicolumn{1}{c|}{17.1698} &
  \multicolumn{1}{c|}{14.7170} &
  \multicolumn{1}{c|}{99.8113} &
  \multicolumn{1}{c|}{36.2264} &
  \multicolumn{1}{c|}{35.0943} &
  \multicolumn{1}{c|}{10.7547} &
  \multicolumn{1}{c|}{25.6604} &
  \multicolumn{1}{c|}{36.6038} &
  20.3774 \\ \cline{2-11} 
\multicolumn{1}{|c|}{\multirow{-2}{*}{\textbf{5x5}}} &
  \multicolumn{1}{c|}{\textbf{$t=T$}} &
  \multicolumn{1}{c|}{46.0377} &
  \multicolumn{1}{c|}{44.3396} &
  \multicolumn{1}{c|}{\textbf{100.0000}} &
  \multicolumn{1}{c|}{87.7358} &
  \multicolumn{1}{c|}{\textbf{89.8113}} &
  \multicolumn{1}{c|}{\textbf{46.9811}} &
  \multicolumn{1}{c|}{57.7358} &
  \multicolumn{1}{c|}{3.2075} &
  15.0943 \\ \hline
\multicolumn{11}{|c|}{\textbf{Baseline (No $\mathcal{L}_{prox}$)}} \\ \hline
\multicolumn{1}{|c|}{} &
  \multicolumn{1}{c|}{$t=\nicefrac{T}{2}$} &
  \multicolumn{1}{c|}{3.6810} &
  \multicolumn{1}{c|}{2.6585} &
  \multicolumn{1}{c|}{3.0675} &
  \multicolumn{1}{c|}{1.0225} &
  \multicolumn{1}{c|}{2.4540} &
  \multicolumn{1}{c|}{1.8868} &
  \multicolumn{1}{c|}{3.0189} &
  \multicolumn{1}{c|}{4.9057} &
  2.6415 \\ \cline{2-11} 
\multicolumn{1}{|c|}{\multirow{-2}{*}{\textbf{1x1}}} &
  \multicolumn{1}{c|}{\textbf{$t=T$}} &
  \multicolumn{1}{c|}{2.8630} &
  \multicolumn{1}{c|}{1.4315} &
  \multicolumn{1}{c|}{3.2720} &
  \multicolumn{1}{c|}{1.6360} &
  \multicolumn{1}{c|}{0.4090} &
  \multicolumn{1}{c|}{2.2642} &
  \multicolumn{1}{c|}{11.3208} &
  \multicolumn{1}{c|}{\cellcolor[HTML]{FFFFFF}6.6038} &
  11.8868 \\ \hline
\multicolumn{1}{|c|}{} &
  \multicolumn{1}{c|}{$t=\nicefrac{T}{2}$} &
  \multicolumn{1}{c|}{11.2474} &
  \multicolumn{1}{c|}{11.0429} &
  \multicolumn{1}{c|}{2.6585} &
  \multicolumn{1}{c|}{3.2720} &
  \multicolumn{1}{c|}{2.8630} &
  \multicolumn{1}{c|}{1.6981} &
  \multicolumn{1}{c|}{9.4340} &
  \multicolumn{1}{c|}{17.1698} &
  13.3962 \\ \cline{2-11} 
\multicolumn{1}{|c|}{\multirow{-2}{*}{\textbf{3x3}}} &
  \multicolumn{1}{c|}{\textbf{$t=T$}} &
  \multicolumn{1}{c|}{18.4049} &
  \multicolumn{1}{c|}{23.1084} &
  \multicolumn{1}{c|}{3.0675} &
  \multicolumn{1}{c|}{1.4315} &
  \multicolumn{1}{c|}{0.8180} &
  \multicolumn{1}{c|}{2.4528} &
  \multicolumn{1}{c|}{43.3962} &
  \multicolumn{1}{c|}{26.2264} &
  46.9811 \\ \hline
\multicolumn{1}{|c|}{} &
  \multicolumn{1}{c|}{$t=\nicefrac{T}{2}$} &
  \multicolumn{1}{c|}{26.3804} &
  \multicolumn{1}{c|}{31.2883} &
  \multicolumn{1}{c|}{3.4765} &
  \multicolumn{1}{c|}{0.2045} &
  \multicolumn{1}{c|}{3.2720} &
  \multicolumn{1}{c|}{2.2642} &
  \multicolumn{1}{c|}{25.2830} &
  \multicolumn{1}{c|}{29.0566} &
  32.6415 \\ \cline{2-11} 
\multicolumn{1}{|c|}{\multirow{-2}{*}{\textbf{5x5}}} &
  \multicolumn{1}{c|}{\textbf{$t=T$}} &
  \multicolumn{1}{c|}{43.7628} &
  \multicolumn{1}{c|}{44.9898} &
  \multicolumn{1}{c|}{3.8855} &
  \multicolumn{1}{c|}{0.2045} &
  \multicolumn{1}{c|}{2.0450} &
  \multicolumn{1}{c|}{2.8302} &
  \multicolumn{1}{c|}{57.5472} &
  \multicolumn{1}{c|}{39.6226} &
  \textbf{60.1887} \\ \hline
\end{tabular}
\caption{Influence of proximity metric and trigger size in the attack against nine robust aggregation schemes. Results are reported as ASR at mid-training and final step. In bold we represent the best attack performance for each RA at the end of training.}
\label{Tab:ScalingBackdoor}
\end{table*}

\subsection{Ablation experiment: Analysing Stealthy Poisoning}

In this experiment, we analyse not only the ASR of our adaptive attacks, but also compare the nature of the attacks stealthiness. For the Krum and Multi-Krum RAS, the aggregator selects the 3 client updates ($m=3$) with the lowest Krum score (explained in Section.\ref{Related Work}). This selection of benign/malicious clients at every training round, can be used as additional metric for a deeper analysis. Specifically, in order to visualise our attack moving into the benign distribution of updates (in terms of the Multi-Krum aggregation rule) we plot the krum score for the 3 benign client updates with the lowest krum score at each training round, for both the baseline attack and our informed attack in CIFAR-10 dataset. 

\begin{figure}[ht]
    \centering
    \includegraphics[width=\linewidth]{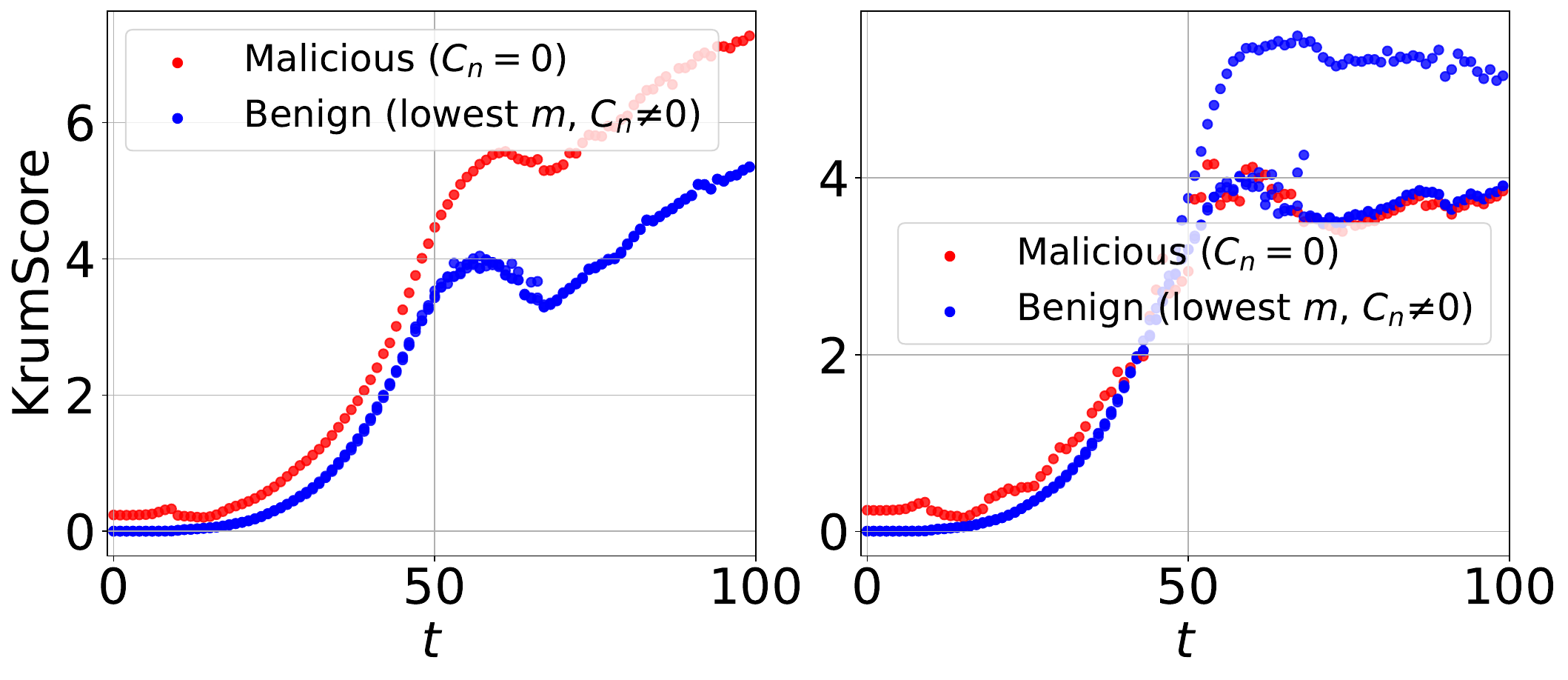}
    \caption{Krum score for a malicious client performing (client 0) a targeted backdoor attack on CIFAR-10, along with the Krum score of the $k$ benign client with the lowest krum scores. Left: Vanilla backdoor attack. Right: Our proposed attack \ourapproach.}
    \label{fig:Krum_score}
\end{figure}

In Fig.\ref{fig:Krum_score}, we can see the aggregator correctly excluding the malicious update from aggregation for the baseline attack, as it allocates a lower krum score to at least three other clients in the FL system. However using the backdoor information leaked by our BSCIl to inform our malicious client, we produce a poisonous update that is able to become one of the $m$ client updates selected by the server for aggregation. 
In Appendix.\ref{APP:B}, we further analyse this behaviour for the other distance metrics to show our attack moving into the distribution of acceptable updates for a range of proximity estimations.

\subsection{Ablation Experiment: Backdoor Side Channel Inference performance}

\subsubsection{Comparative measurements for Backdoor inference} 
In this experiment, we analyse how the global model leaks side channel information on the membership of a backdoor being implemented into the global model. Moreover, we compare the use of this BSCI estimation to balance the loss function of our chameleon attack versus a more direct alternative consisting on simply measuring the ASR of the malicious clients poisoned samples on the published global model at each iteration. This alternative can be expressed mathematically as:
\begin{equation}
         \alpha_t = 1 - (\sum_{i = t - k}^t\frac{ASR_i}{k})
    \label{eq:alpha2}
\end{equation}

Our hypothesis is that BSCI, while more complex, is able to detect earlier signs of poisoning, not yet visible in the class estimation ASR, that can be used as early guide for our adaptive attack. Furthermore, that since it allocates a likelihood to the global model being backdoored, it is able to implement stronger poisoning once it is in the distribution.

In Fig.\ref{fig:infplot}, we analyse the resultant calculation of BSCI inference success $v_t$ at every training round. For this comparison, we plot the inference success in blue, ASR in orange, along with the membership of the malicious model made it into the set of acceptable clients for aggregation in red. We repeat the analysis for defences Krum, Multi-Krum, Bulyan and DAI against CIFAR-10.

\begin{figure*}[h]
    \centering
    \includegraphics[width=\textwidth]{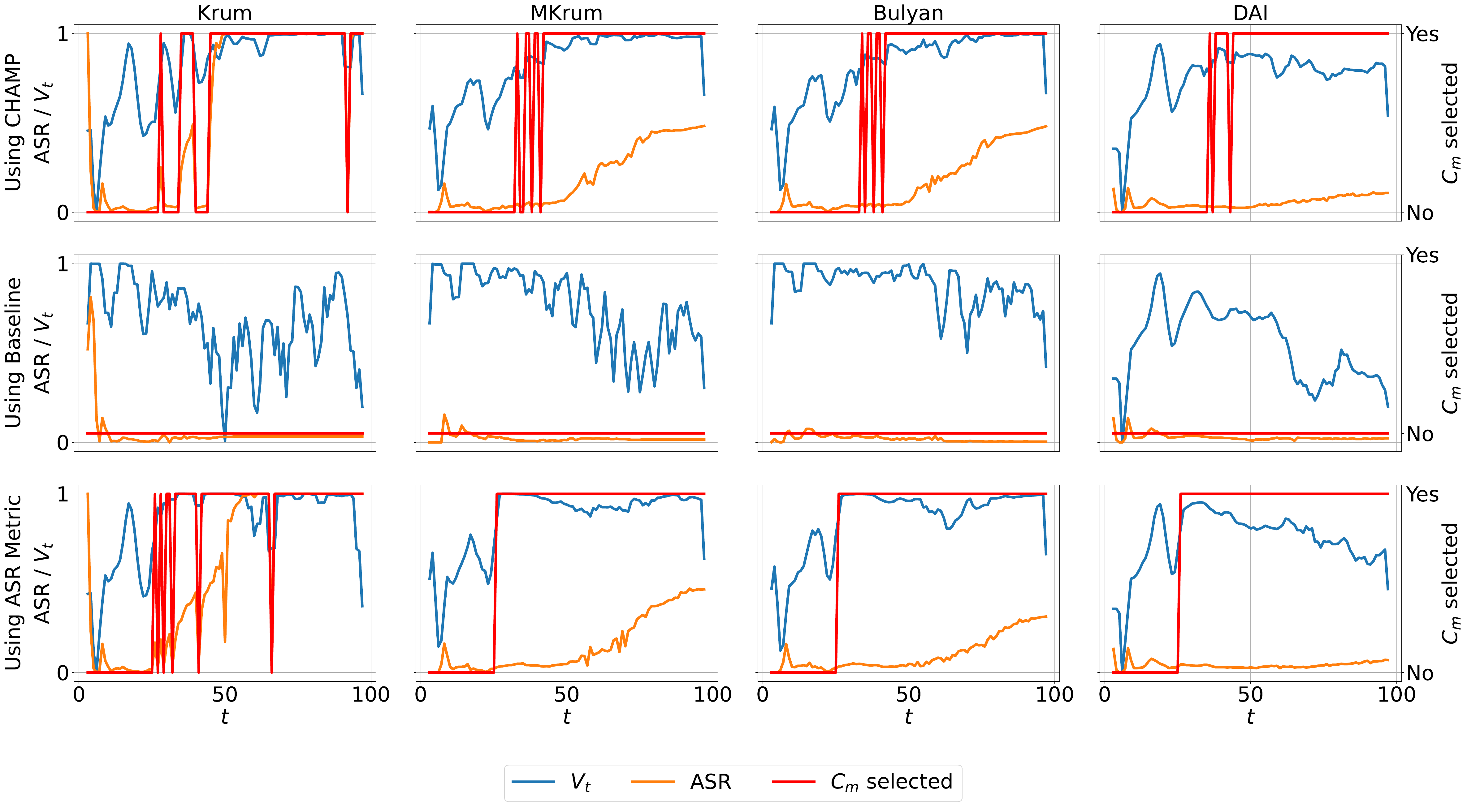}
    \caption{Figure showing both the backdoor side channel information leakage value $v_t$, ASR in a range between 1 and 0 for readability, and whether the model has been selected by an aggregation scheme, for a single pixel backdoor attack on Cifar-10 in different settings.}
    \label{fig:infplot}
\end{figure*}

Fig.\ref{fig:infplot} shows the utility of \ourapproach over data poisoning and also the need for our proposed BSCI inference model instead of simply measuring the utility of the attack by measuring ASR for the published global model across backdoored samples in the local client. Comparing the first two rows we can see how our inference model over time predicts the global model as more likely to be non-backdoored as training rounds increase. Using \ourapproach, the malicious actor is able to catch this behaviour and start altering it's $\mathcal{L}_{pois}$ to implement poisoning, thus managing to increase ASR that revealed the attack success. Furthermore, note that in the plots using \ourapproach, our model update toggles between being inside and outside the distribution of accepted updates. When this happens, our inference model is able to detect this behaviour and $v_t$ is lowered, therefore making our attack more stealthy.

Investigating \ourapproach in comparison to using ASR as a proximity metric from Fig.\ref{fig:infplot}, the utility of the inference model made clear. Using a measure of proximity such as ASR over backdoored samples, produces an update that can also circumvent the RA. However, since ASR is low when no attack makes the distribution, this attack estimates that it needs to produce an update that relies heavily on decreasing the distance to the global model, therefore it produces an update that apparently gets into the distribution in comparatively few training rounds. However, once it gets into the distribution there is no immediate increase in ASR, as there is in our inference model prediction (see Multi-Krum using ASR as a proximity metric plot for a good example of this). So the update continues to focus more on being in distribution than poisoning, whereas our inference model has access to more fine-grained information, and therefore is able to produce updates that once in the distribution, implement the maximal amount of poisoning possible.

\subsubsection{$\alpha$ convergence}

Following from the previous experiment, in this experiment, we aim to show that, using our backdoor side channel, we can converge to a $\alpha$ value that is optimal for poisoning. To analyse this we present the alpha values calculated using Equation.\ref{eq:alpha} for differing system setups.
\begin{figure*}[h]
    \centering
    \includegraphics[width=\linewidth]{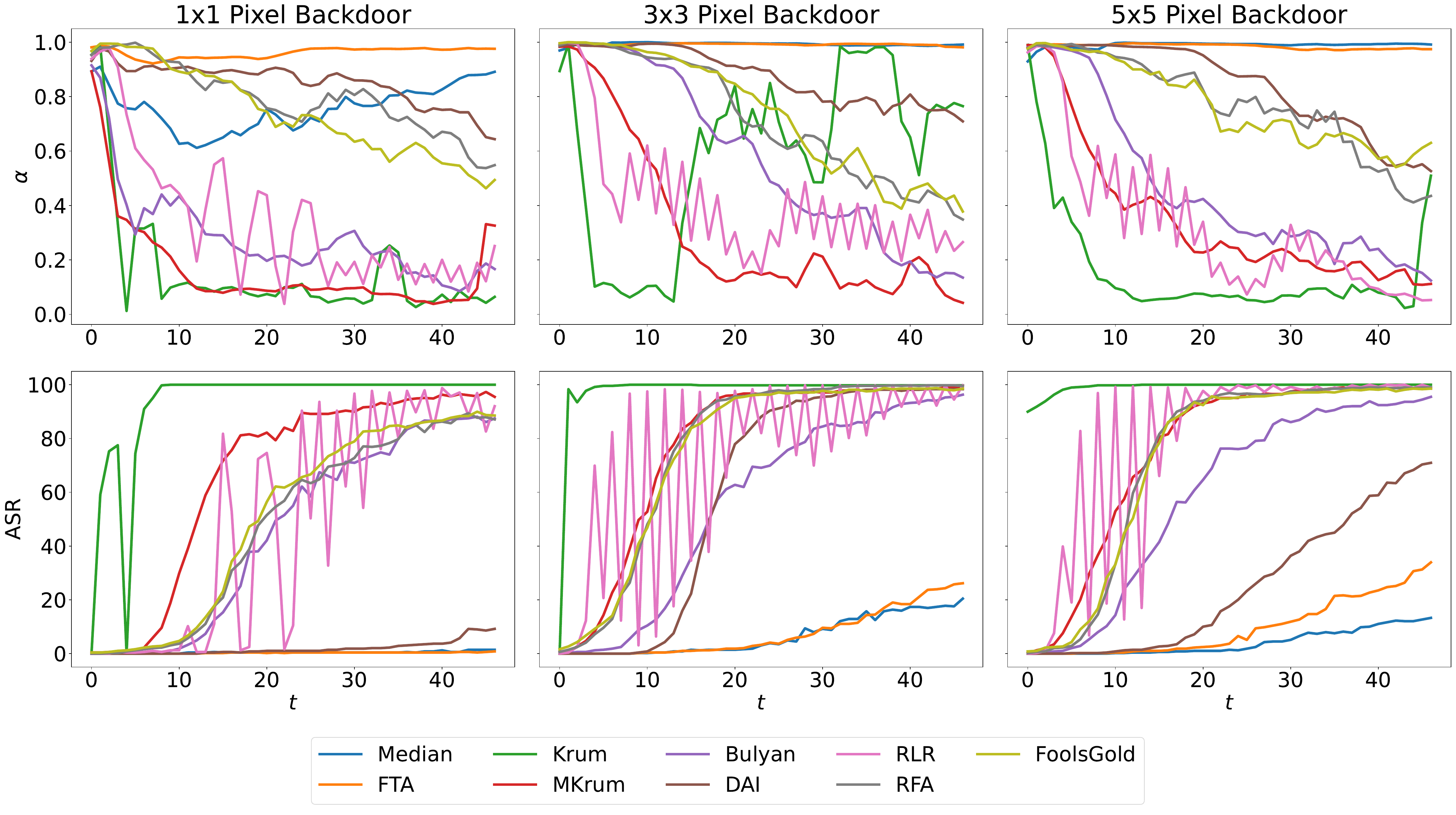}
    \caption{Figure showing the $\alpha$ score and ASR calculated for each RA scheme at every training round using Eq.~\ref{eq:alpha} starting from training round $k$ for Fashion-MNIST.
    }
    \label{Alphas}
\end{figure*}

In Figure.\ref{Alphas} we can see how our $\alpha$ parameter is related to the ASR of our attack against each defence. We can see that the $\alpha$ values that the malicious client calculates throughout training is dependent on the system settings such as the RA being used and the backdoor being implemented. We can see some general trends emerging such as Median having a high $\alpha$ value and Krum having a low value for two out of three backdoors (one we break the Krum aggregation scheme the malicious update becomes the new global model, so it is intuitive that we do not have to change much from that point), but there is no intuitive way to arrive at these $\alpha$ values without an adaptive attack that infers side channel information at each training round. Highlighting the utility of our attack in finding a value to implement maximal poisoning into the system.

\section{Related Work}\label{Related Work}
\noindent\textbf{Model Poisoning Attacks against RA schemes}:
Multiple model poisoning and backdoor attacks have been proposed that try to circumvent RA schemes under varying assumptions and system setups. In \cite{bagdasaryan_how_2020}, the constrain-and-scale attack is introduced to constrain the loss function of malicious clients updates such that the attacker can circumvent anomaly detection schemes. 

In \cite{fang_local_2020}, untargeted poisoning attacks were formulated to attack specific robust aggregation schemes in FL. Similarly in \cite{shejwalkar_manipulating_2021}, a comprehensive study of untargeted model poisoning attacks was conducted under differing threat models. It was shown that these attacks are effective in increasing the ASR of attacks against the selected RA schemes, and also when attacking schemes that the attack was not optimised for. Further papers \cite{wang_attack_2020,sun_can_2019, zhou_deep_2021,hossain_desmp_2021,fang_local_2020} discuss model poisoning in differing settings (rare samples, against differential privacy etc.). We propose an attack against a stronger threat model as discussed in Sec.\ref{OurAttack}.

In \ourapproach we move afar from these papers and aim to not only circumvent a RA defence, but to implement an attack in a black box setting that implements the \textit{greatest possible poisoning} into the global model, all whilst remaining camouflaged (unseen by the RA). Using our inference, we converge to a value of $\alpha$ that implements this poisoning. Our approach is also evaluated against, not a selected RA group, but an extensive set of state-of-art RA schemes.

\textbf{Membership inference in FL}:
In FL,  MIAs are commonly launched by a central server in a server-side attack. This takes place when a malicious server performs membership inference attacks on local client updates \cite{nasr_comprehensive_2020}. These attacks perform well as the impact of the local client data should be easier to detect in the model update, as the local model has been re-tuned on local data since it has been aggregated. However since local training still starts from an aggregated model, they exhibit a decrease in performance when compared to attacking centrally trained models. Moreover, they require a malicious actor to capture the server.

A malicious client can perform client-side attacks on the global model in order to infer the presence of a sample in another client's dataset. One method for this is gradient ascent, where the malicious client aims to increase the loss over a specific sample. If this sample exists inside the local data-set of another client, the loss over this sample will remain low, revealing the presence of this sample in another client's dataset to the attacker. In \cite{nasr_comprehensive_2020}, both server and client side white box inference attacks in FL are analysed. It is shown that FL is inherently open to white-box attacks as the global model is shared across the system to all clients. Furthermore, it is shown that increasing the number of clients in a FL system decreases the effectiveness of client-side MIAs. 

As main differentiator, our approach \ourapproach uses membership inference principles however differs in what information we aim to gain. The adversary launches a membership inference estimation on its own dataset, to infer the presence of specific poisoned data points in an aggregated model, which is used as side-channel feedback to measure its own success. Furthermore, much like the observation on affecting the loss of specific samples in the gradient ascent approach, we observe that outlier samples (here backdoored samples) are easier to detect by a membership inference model.

\noindent \textbf{Robust Aggregation Schemes.}
RA schemes attempt to remove outliers, which are equated to malicious behaviour, to the standard distribution of updates. Below, we outline a few key RA schemes we wish to target: 

\textbf{Median \cite{yin_byzantine-robust_2018}}: Defined as the co-ordinate by co-ordinate median of every uploaded model to the central server. We define this scheme for a co-ordinate $k \in K$ as 
\begin{equation}
    Median \{g^t_{nk}:n\in N\}=[median(g^t_{1,k},…,g^t_{N,k})]        
\end{equation}

\textbf{Trimmed Mean \cite{yin_byzantine-robust_2018}}: Defined as trimmed mean of every co-ordinate for every uploaded model to the central server. For every co-ordinate across a set of uploaded models to the central server, we remove the maximum and minimum $\beta$ values and take the mean of the remaining values as the global models parameter. So for every parameter $k \in K$, we have a set of parameters that are inside of the most extreme values defined as $S^t_k$, then we can say for each co-ordinate $k$.
\begin{equation}
\begin{split}
        \text{Trimmed-Mean}\{g_{nk}^t:n\in N\} = \big[& \frac{1}{N-2\beta N}\sum_{g^t_{n,1} \in S^t_1} g^t_{n,1}]
\end{split}
\end{equation}

\textbf{Krum \cite{blanchard_machine_2017}}: For each client update at training round $t$, the krum algorithm computes the distance between the client update and all other client updates using squared euclidean distance. A krum score is then assigned to each client that is the sum of distances to the nearest $N-f-2$ other updates, where $f$ is the number of byzantine clients assumed by the central server. The client with the lowest Krum score is selected as the new global model. So for each client, the server calculates the krum-score:
\begin{equation}
    \text{Krum-score}(C_n) = \sum_{j \in \hat{N}}||g^t_n - g^t_j||_2^2
\end{equation}
where $\hat{N}$ is the set of $N-f-2$ clients closest to $n$ in terms of euclidean distance. Krum finds the client model with the minimum Krum-score (we define as $g^t_*$, and the global model is set to:
\begin{equation}
    \text{Krum}\{g_{n}^t:n\in N\} =g^t_*
\end{equation}

\textbf{Multi-Krum \cite{blanchard_machine_2017}}: A variation on Krum, where isntead of selecting one model as the new global model, we select $m$ clients with the lowest Krum score and average them as the new global model. Once the Krum-score is calculated then, we have a set of $m$ clients in our set $g^t_*$, which we simply average over:

\begin{equation}
    \text{Multi-Krum}\{g_n^t:n\in N\} = \text{FedAvg}\{g_{*}^t\}
\end{equation}

\textbf{Bulyan \cite{mhamdi_hidden_2018}}: Bulyan combines multi-krum with a co-ordinate by co-ordinate robust aggregation scheme. That is, once $N-2f$ clients are selected by utilising multi-krum, the aggregator performs a co-ordinate wise robust aggregation scheme such as median of trimmed-mean across the selected model updates. So once we found our set of client models $g^t_*$, we do:
\begin{equation}
    \text{Bulyan}\{g_n^t:n\in N\} = \text{Trimmed-mean}\{g_{*}^t\}
\end{equation}

\textbf{RFA \cite{pillutla_robust_2022}}: Robust federated aggregation, or RFA, iteratively computes the geometric median of client updates using Weiszfeld's algorithm. This algorithm iteratively finds a vector that minimizes the Euclidean distance to a set of vectors, here, the local model updates.A parameter $\epsilon$ is used as stability constant for smoothing distances to client models. It finds a global model $G^t$, such that:
\begin{equation}
    G^t = arg\min_{k \in K}\sum^N_{n=1} ||G-g^t_n||_2
\end{equation}
where $G$ is an approximation of $G^T$ during the execution of the Weiszfeld's algorithm. This algorithm will continue to iterate until either the difference between two iterations is deemed small enough ($tol$) or enough iterations have passed. 

\textbf{AlignIns \cite{xu_detecting_2025}}: AlignIns performs direction alignment inspection using cosine similarity. In this scheme, the previous rounds global model is taken as a reference model for the benign direction of convergence. Each update is then given a direction alignment score using cosine similarity based on it's alignment to the benign direction. For each client, we will calculate the average cosine similarity $\alpha_n$ of the direction alignment $u^t_n = \nicefrac{g^t_n}{||g^t_n||_2}$ of each client. Then, the clients that fall within a threshold -similar to the $\beta$ parameter in the Trimmed-Mean aggregation scheme- are selected as the set of allowed client updates $g^t_*$. Then we simply use FedAvg to aggregate the allowed updates:

\begin{equation}
    \text{DAI}\{g_n^t:n\in N\} = \text{FedAvg}\{g_{*}^t\}
\end{equation}


\textbf{RLR \cite{ozdayi_defending_2021}}: In the Robust Learning Rate (RLR) defence, the central server applies a scalar value to each received before aggregation. This scalar value $\lambda_n$ is calculated by taking the agreement of all clients on whether a client update is pushing the global model in the correct direction. Through this, clients with highly dissimilar updates are scaled out of the aggregation ($\lambda_n \to 0 $). The calculation of $\lambda_n$ is done using parameters $c,\theta$ which control scaling and the threshold for boundary agreement decisions. Final aggregation is obtained such that:
\begin{equation}
    RLR\{g_{nk}^t:n\in N\} = \big[ \frac{1}{N}\sum_{n=1}^N \lambda_ng^t_{n,1}]
\end{equation}

\textbf{FoolsGold \cite{fung_mitigating_2020}}: Similar to RLR, FoolsGold applies a scalar value to each client before aggregation. Cosine similarity is used to punish updates that attempt to push the global model in a malicious direction. However this defence is aimed at stopping sybil attacks, that is, colluding attacks against a FL system. 
\begin{equation}
    FoolsGold\{g_{nk}^t:n\in N\} = \big[ \frac{1}{N}\sum_{n=1}^N \lambda_ng^t_{n,1}]
\end{equation}


\section{Concluding Remarks}\label{Conclusion}
In this paper, we propose a novel adaptive poisoning attack framework, Chameleon Poisoning (CHAMP), that exploits side-channel feedback from
the aggregation process to guide a stealthy and effective attack. Our approach is able to infer using side channel feedback from the global model to reliably steer and balance the malicious updates in-distribution without white-box
access to the server or its aggregation rule.

Our approach is evaluated in two different datasets against nine state-of-art RA schemes, being able to consistently improve the attack success rate of conventional backdoor attacks. More importantly, we demonstrate that it is possible, under a suitable configuration, to break all RA defences and introduce a non-negligible amount of poisoning, similar to the performance under no RA scheme or above. 

Therefore, this framework challenges the prevalent assumption malicious updates are inherently out-of-distribution, which forms the base of all existing RA schemes. By highlighting this fundamental limitation of existing RA defenses, we underscore the need for new strategies to secure federated learning against sophisticated adversaries.

As future work, we aim to propose a novel RA scheme that is able to counter-attack to the proposed framework in this paper. We also aim to exploit the BSCI estimation to distinguish benign and malicious out-of-distribution updates, allowing RA algorithms to be more fair while still robust against attacks.

\bibliographystyle{IEEEtran}
\bibliography{references, bib}
\newpage
\appendices

\section{Membership inference over backdoored samples}\label{APP:A}
In \cite{shokri_membership_2017}, the membership inference accuracy over the MNIST dataset is similar to random chance. This is due to the similarity of hand-drawn digits in the data-set. We aim to prove that if members of the data-set are backdoored, then this change in the distribution of data has a greater level of data leakage than benign samples, and can therefore be exploited by an inference model. 

To do this analysis, we train a model using the model architecture outlined in \cite{mcmahan_communication-efficient_2017}, for 10 training rounds using centralised learning on the MNIST data-set. We then train two shadow models, both of which have been backdoored across all samples of the source class, which is set to the digit 0. For the sake of this experiment we do not change the labels of these samples, as we are simply interested in how this backdoor changes the distribution across these samples and therefore leaks more information to the attacker. These shadow models are used to train a classifier to perform a membership inference attack using AIJACK\footnote{Available at \hyperlink{https://koukyosyumei.github.io/AIJack/index.html}{https://koukyosyumei.github.io/AIJack/index.html}}, against a fully backdoored global model. The MIA model is then ran against the output of the trained target model across backdoored source class samples and non-backdoored source class samples. For comparison, we do the same experiment against target models which contain no backdoored samples, using shadow models that have also not been backdoored. The results of this are seen in Fig.\ref{fig:CMS} and Fig.\ref{fig:roc}.

\begin{figure}
    \centering
    \includegraphics[width=\linewidth]{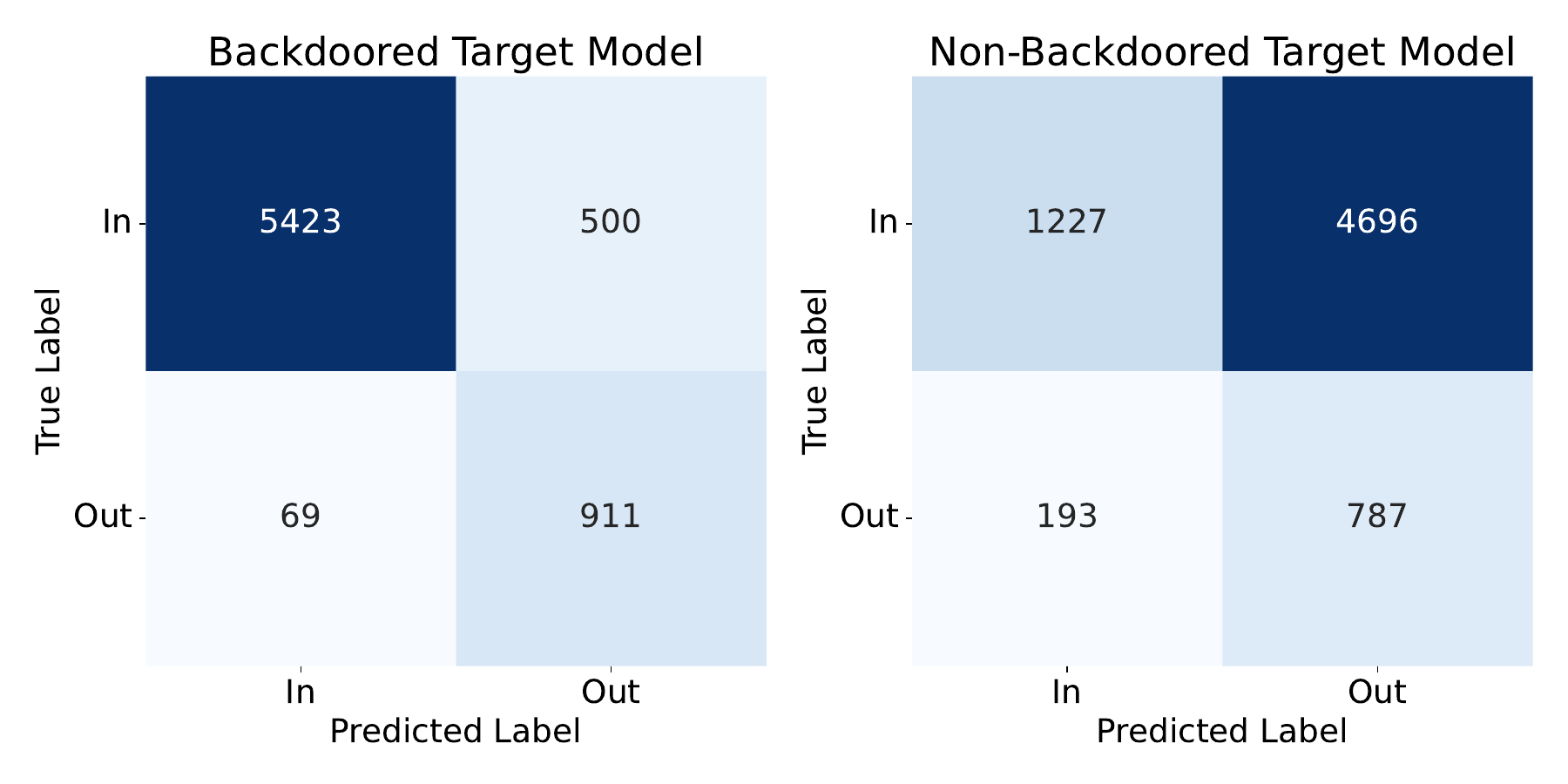}
    \caption{Confusion matrix for MIAs against backdoored and non-backdoored target models.}
    \label{fig:CMS}
\end{figure}

\begin{figure}
    \centering
    \includegraphics[width=\linewidth]{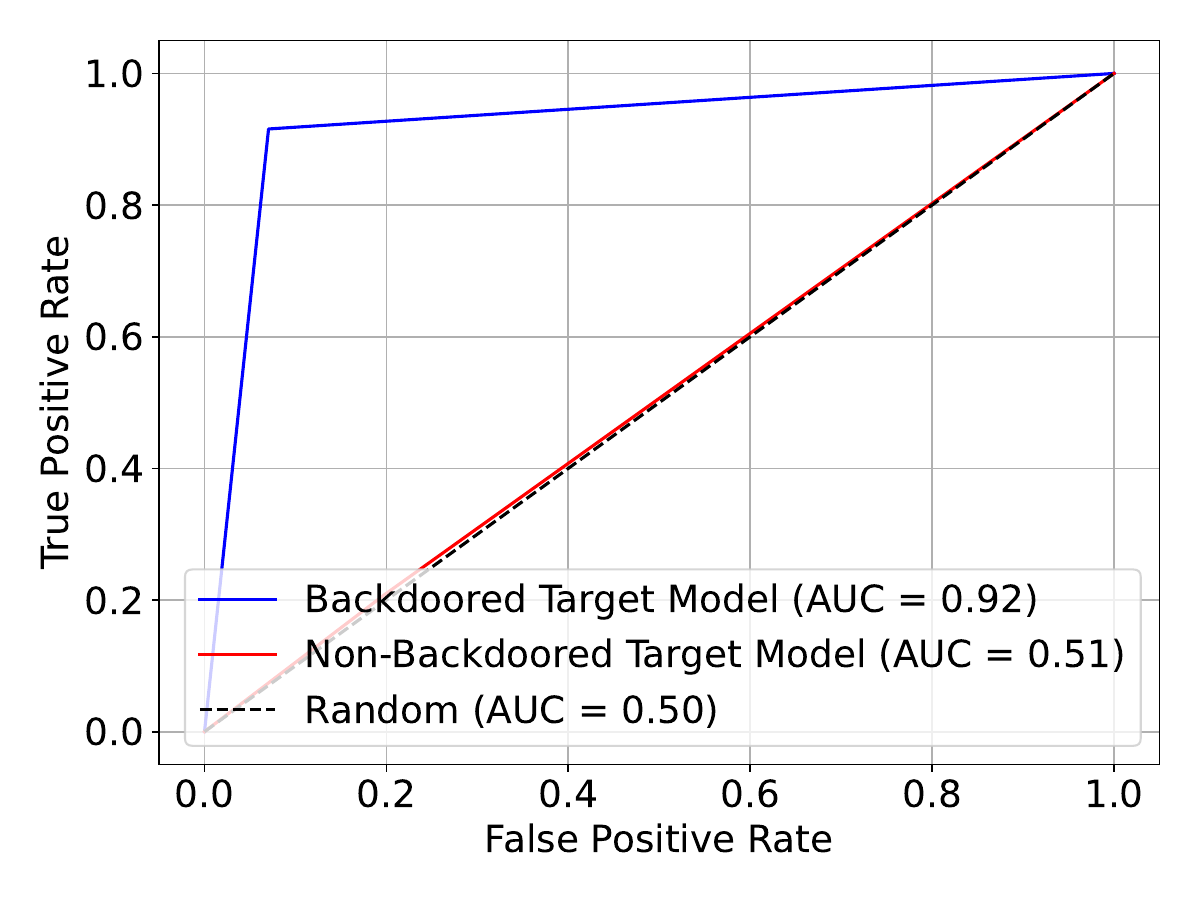}
    \caption{ROC curve for MIAs against backdoored and non-backdoored target models.}
    \label{fig:roc}
\end{figure}

We conclude experimentally, that altering the data such that a backdoor is seen across the set of samples that are considered inside the training data-set (\enquote{in}) by the MIA, and not in the training data-set (\enquote{out}), improves the attack success of the MIA as this distribution of data is easier to identify than benign data samples.

This analysis proves that backdoored behaviour is easier to identify in a target model. We use this in \ourapproach, as a basis for our attack as malicious samples should be easier to identify in the global model using membership inference than benign samples. Therefore we should be able to detect them in a model trained using FL, allowing our attack to work and overcoming the observation made in \cite{nasr_comprehensive_2020}, that aggregation hides the data leakage from specific samples in the global model in a FL system.

\section{Poisoned model distance during training}\label{APP:B}
For a targeted single pixel backdoor attack against the multi-krum aggregation scheme, we plot the Euclidean distance and cosine similarity of the malicious client and the 3 clients with the lowest metric score, along with a TSNE plot of the model updates at the final training round for both the data poisoning attack and our model poisoning attack.

\begin{figure}[htbp]
    \centering
    \includegraphics[width=\linewidth]{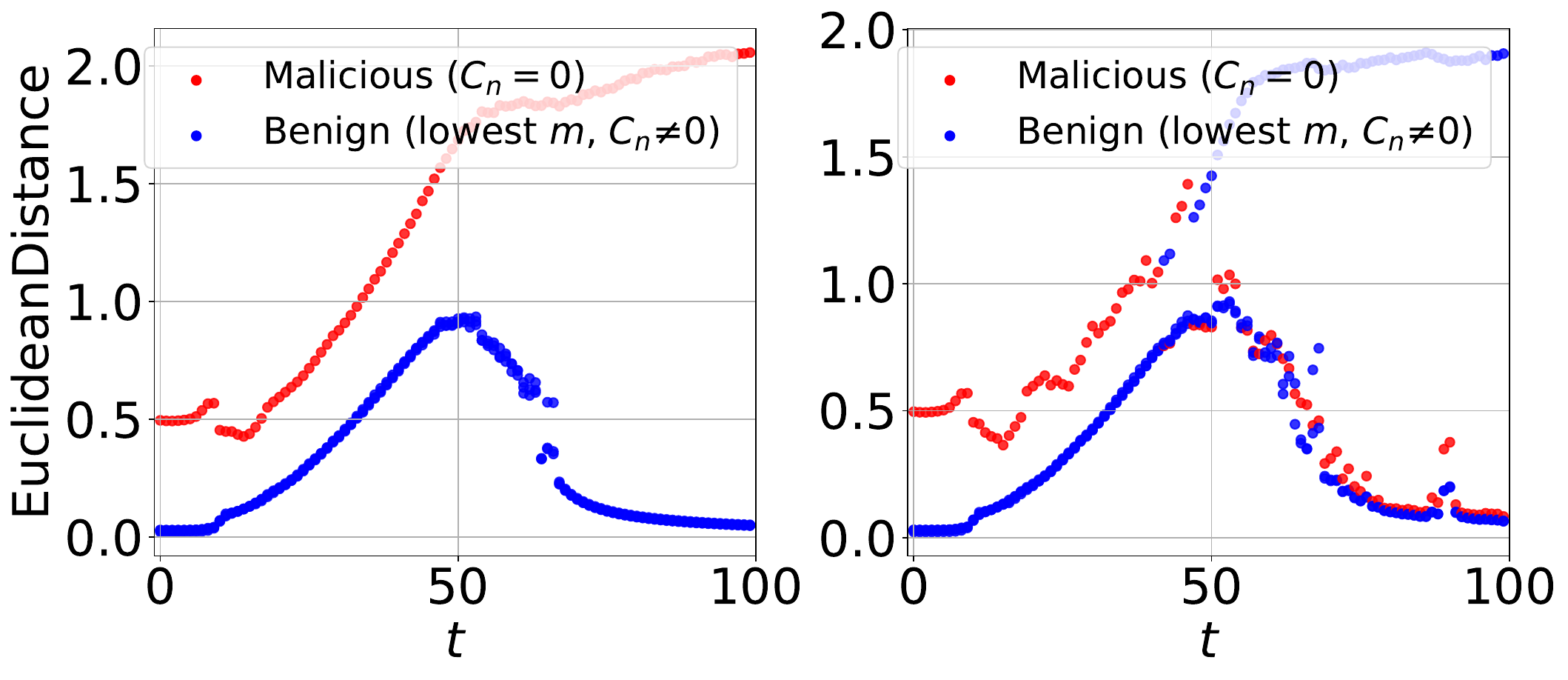}

    \vspace{1em} 
    \includegraphics[width=\linewidth]{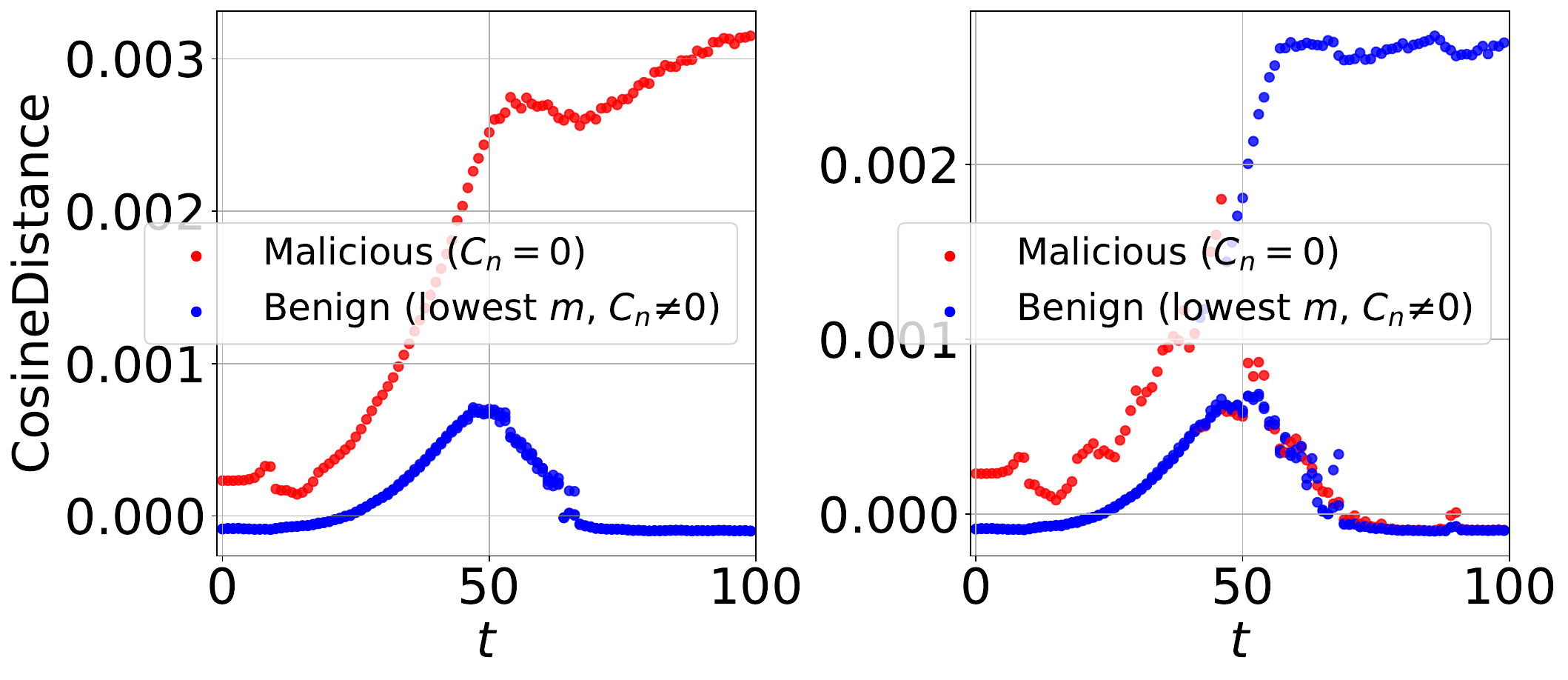}
    \caption{Visualization of the classic backdoor attack (LHS) and our attack (RHS) in terms of euclidean distance to other client updates in the aggregation.}
    \label{fig:four_plots}
\end{figure}
In Fig.\ref{fig:four_plots} we can see our model updates move closer to the aggregation, and we know from Section.\ref{Experimental Results}, that it keeps it's poisonous behaviour as well as being close in proximity to other client's updates. As further evidence, we plot a TSNE plot for both experiments, of all final model update parameters to the server (at round $t=T$) in Fig.\ref{fig:tsne}. As we can see our attack is closer to and embedded into other updates and does not lie at the edges of the plot as we see in the classic backdoor attack.

\begin{figure}[htbp]
    \centering
    \begin{subfigure}[t]{0.48\linewidth}
        \centering
        \includegraphics[width=\linewidth]{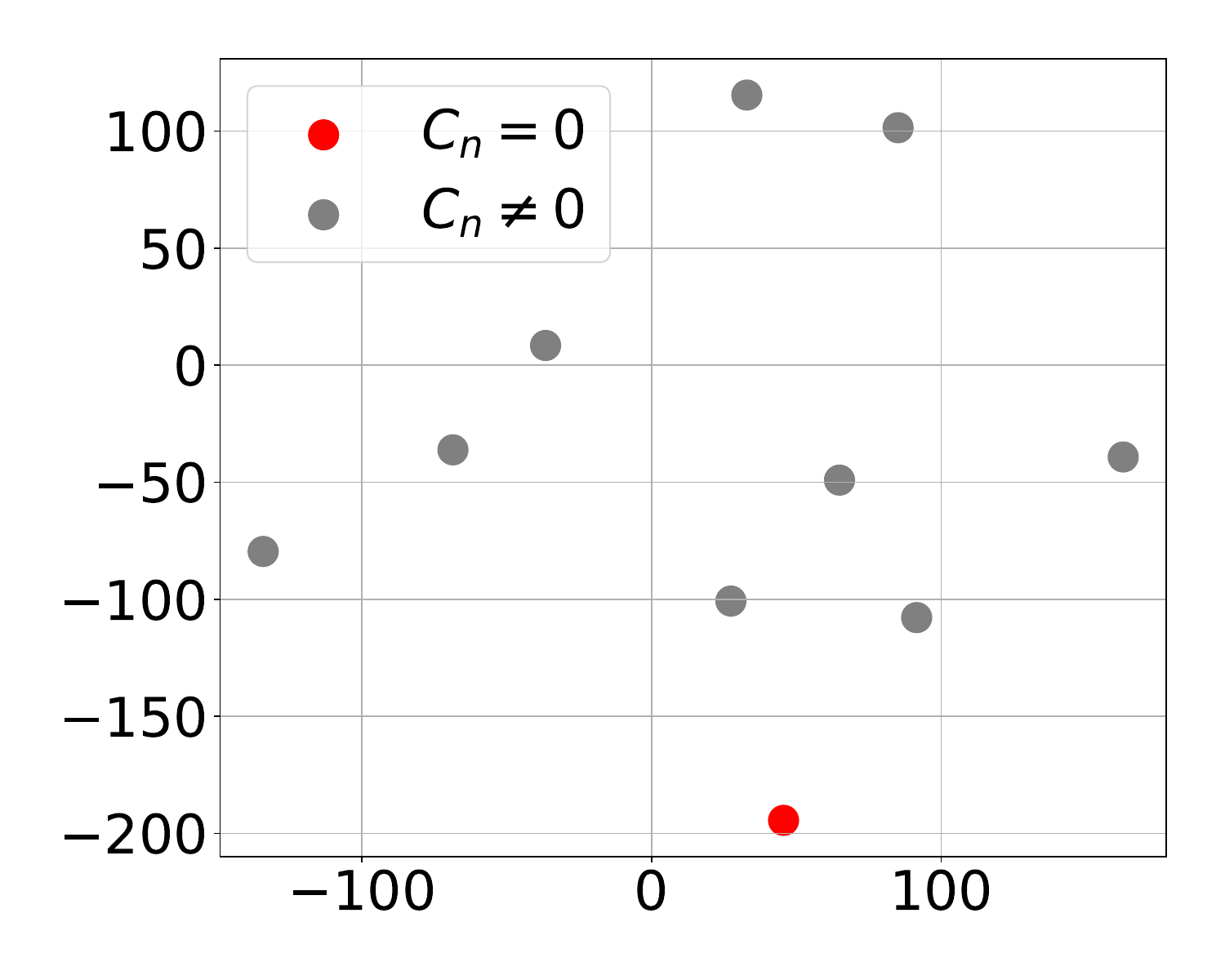}
        \label{fig:plot1}
    \end{subfigure}
    \hfill
    \begin{subfigure}[t]{0.48\linewidth}
        \centering
        \includegraphics[width=\linewidth]{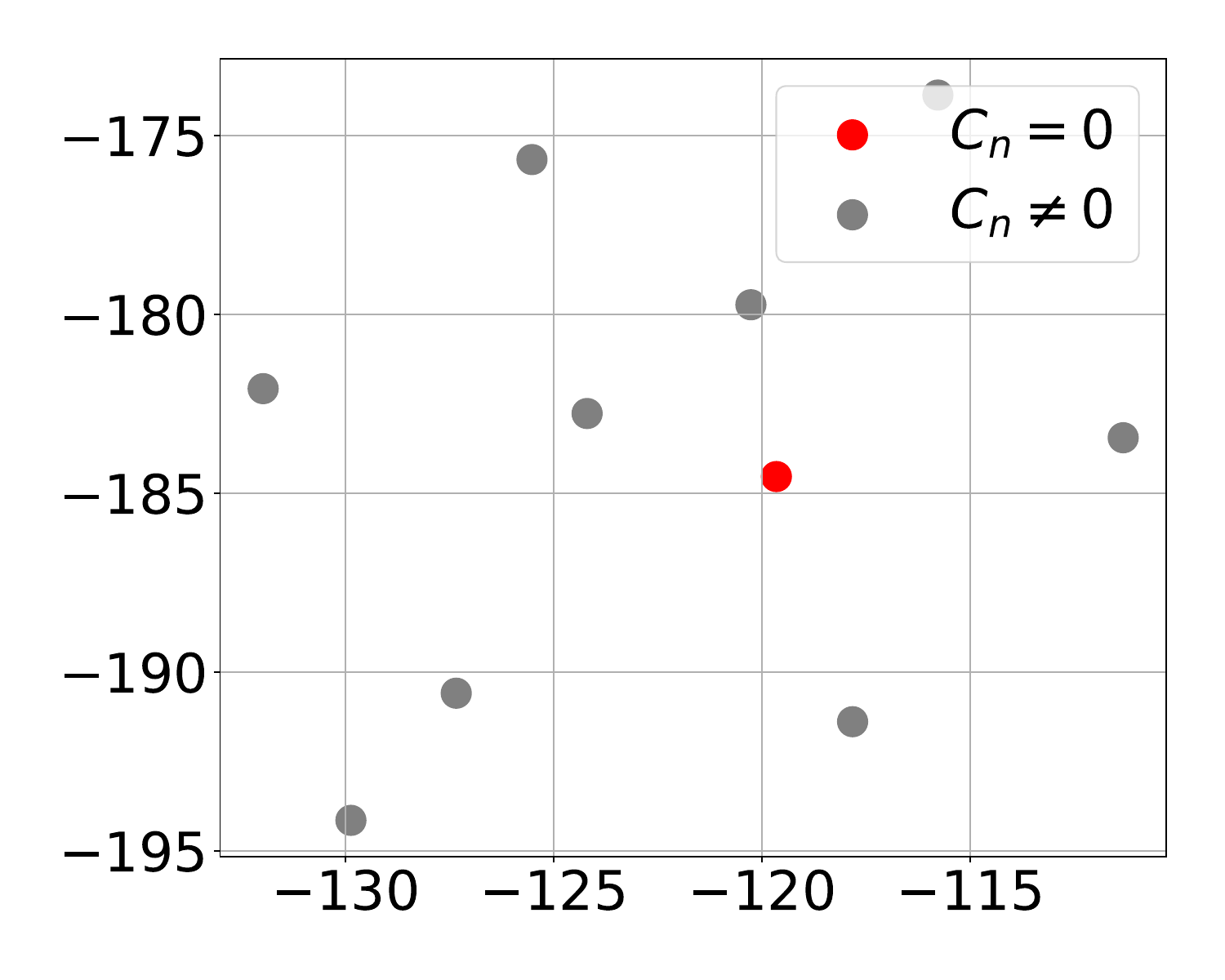}
        \label{fig:plot2}
    \end{subfigure}
    \caption{TSNE plots for all clients updates to a central server at training round $T$. For a FL system that has been under baseline attack (left) and attacked using \ourapproach (right).}
    \label{fig:tsne}
\end{figure}
\section{Model Architectures}\label{APP:C}
In Table.\ref{tab:FMNISTCNN} and Table.\ref{tab:CifarCNN}, we outline the model architectures used for both data-sets investigated in this paper.
\begin{table}[ht]
\centering
\caption{CNN architecture used for Fashion-MNIST classification.}
\begin{tabular}{l|l|c}
\hline
\textbf{Layer} & \textbf{Configuration} & \textbf{Output shape} \\
\hline
Input           & Grayscale image & $1 \times 28 \times 28$ \\
Conv1 + ReLU    & 30 filters, $3 \times 3$, padding=1 & $30 \times 28 \times 28$ \\
MaxPool         & $2 \times 2$, stride=2 & $30 \times 14 \times 14$ \\
Conv2 + ReLU    & 50 filters, $3 \times 3$, padding=1 & $50 \times 14 \times 14$ \\
MaxPool         & $2 \times 2$, stride=2 & $50 \times 7 \times 7$ \\
Flatten         & -- & 2450 \\
Fully Connected 1 & 100 hidden units, ReLU & 100 \\
Fully Connected 2 & 10 output units (logits) & 10 \\
\hline
\end{tabular}
\label{tab:FMNISTCNN}
\end{table}

\begin{table}[ht]
\centering
\caption{AlexNet CNN architecture for CIFAR-10 classification.}
\resizebox{\linewidth}{!}{%
\begin{tabular}{l|l|c}
\hline
\textbf{Layer} & \textbf{Configuration} & \textbf{Output shape} \\
\hline
Input             & RGB image & $3 \times 32 \times 32$ \\
Conv1 + ReLU      & 64 filters, $11 \times 11$, stride=4, padding=5 & $64 \times 8 \times 8$ \\
MaxPool           & $2 \times 2$, stride=2 & $64 \times 4 \times 4$ \\
Conv2 + ReLU      & 192 filters, $5 \times 5$, padding=2 & $192 \times 4 \times 4$ \\
MaxPool           & $2 \times 2$, stride=2 & $192 \times 2 \times 2$ \\
Conv3 + ReLU      & 384 filters, $3 \times 3$, padding=1 & $384 \times 2 \times 2$ \\
Conv4 + ReLU      & 256 filters, $3 \times 3$, padding=1 & $256 \times 2 \times 2$ \\
Conv5 + ReLU      & 256 filters, $3 \times 3$, padding=1 & $256 \times 2 \times 2$ \\
MaxPool           & $2 \times 2$, stride=2 & $256 \times 1 \times 1$ \\
Flatten           & -- & 256 \\
Fully Connected   & Linear, 256 $\rightarrow$ 10 & 10 \\
\hline
\end{tabular}%
\label{tab:CifarCNN}
}
\end{table}

\end{document}